\documentclass[12pt]{article}
\usepackage{amssymb,amsmath,epsfig}

\begin{document}
\title{\bf Greybody Factor for a Non Accelerated Charged Modified Black Hole in anti-de
Sitter Regime}
\author{M. Sharif \thanks{msharif.math@pu.edu.pk} and A. Raza
\thanks{aliraza008.math@gmail.com}\\
Department of Mathematics and Statistics, The University of Lahore,\\
1-KM Defence Road Lahore, Pakistan.}
\date{}
\maketitle

\begin{abstract}
This paper investigates the greybody factor for a non accelerated
black hole with modified Maxwell electrodynamics in an anti-de
Sitter regime. For this purpose, we compute the radial equation for
a massless scalar field with the help of Klein-Gordon equation. We
then formulate effective potential by transforming this equation
into Schrodinger wave equation. We analyze the graphical behavior of
effective potential for different values of mass, parameter
characterizing the modified Maxwell theory, anti-de Sitter radius
and electromagnetic charge parameters. The exact solutions are
computed at two different horizons, i.e., event and cosmological
horizons through the radial equation. Furthermore, we match the
obtained solutions in an intermediate regime to enhance feasibility
of the greybody factor over the entire domain and check its behavior
graphically. It is found that the greybody factor has a direct
relation with the radius, electromagnetic charge as well as angular
momentum of the black hole and an inverse relation with the anti-de
Sitter radius and modification parameter. We conclude that the
modified Maxwell solution reduces the emission rate of the black
hole.
\end{abstract}

\section{Introduction}

The mysterious features of black holes (BHs) have attracted the
attention of many researchers in recent years. It has such a strong
gravity that prevents all kinds of outgoing material as well as
electromagnetic radiations and absorbs all the surrounding material.
The boundary around the BH is known as event horizon (EH) beyond
which the escape velocity exceeds the speed of light. Later, it was
discovered that BH could emit radiations (Hawking radiations) due to
quantum effects \cite{1}. The mass of BH decreases due to the
emission of radiations which leads to the death of BH. Black holes
are thermal objects (thermodynamics laws hold) and display thermal
properties like Hawking temperature and entropy which are different
for distinct BHs \cite{2}-\cite{4}. The emission rate of Hawking
radiations is defined as \cite{1}
\begin{equation}\nonumber
\Theta(w)=\left(\frac{(2\pi)^{-3}d^3\kappa}{e^{\frac{w}{t_{h}}}\pm
1}\right),
\end{equation}
where $w,~d^3\kappa$ and $t_{h}$ represent frequency, change in
surface gravity and temperature of BH, respectively. The
positive/negative sign shows fermion/boson particles and the above
expression can also be used for higher dimensions.

Hawking radiation is a theoretical prediction made by physicist
Stephen Hawking in 1974. It suggests that BHs are not completely
black but instead emit radiation due to quantum effects near the EH.
This phenomenon arises from the interplay between quantum mechanics
and GR. The outer region of the EH plays a role in modifying the
emitted Hawking radiation. As particles are created and annihilated
near the EH, they can escape the BH due to quantum effects,
resulting in the emission of radiation. However, the curvature of
spacetime caused by the strong gravitational field of the BH can
affect this radiation emission. The concept of the EH acting as a
barrier is important. Particles near the EH are influenced by the
strong gravitational field and some can gain enough energy to escape
the BH gravitational pull as Hawking radiation. However, particles
that are very close to the EH can be captured by the BH
gravitational pull before they escape, essentially creating a kind
of gravitational barrier. This interaction between the particles and
the gravitational barrier influences the radiation's energy
spectrum. The emission rate for a distant observer is expressed as
\begin{equation}\nonumber
\Theta(w)=\left(\frac{|G_{w,l,m}|^{2}d^3\kappa}{(2\pi)^{3}(e^{\frac{w}{t_{h}}}\pm
1)}\right),
\end{equation}
where $|G_{w,l,m}|^{2}$ displays the greybody factor (GBF) which is
a frequency-dependent quantity. The GBF (rate of emission of
particles) is defined as the probability of absorption of a wave by
BH coming from infinity \cite{5}. These waves provide information
about the interior of BH in the form of charge, mass and angular
momentum.

The radiation emitted by a BH, if the effects of the outer curvature
were ignored, would have the characteristics of a black body
spectrum, which is a characteristic spectrum for an idealized body
that absorbs all radiations incident upon it. However, when
considering the effects of the outer curvature and the gravitational
barrier, the emitted radiation spectrum gets modified. This modified
spectrum is often referred to as a greybody spectrum because it is a
combination of the black body spectrum and the modifications caused
by the curvature of spacetime near the event horizon. Thus, the
outer curvature of the EH impacts the emitted Hawking radiation,
causing it to deviate from a black body spectrum. This deviation
results in greybody spectrum, where the radiation appears to be a
mixture of black body and modified emissions due to the
gravitational effects near the event horizon.

In the context of GR, singularity refers to a point in spacetime
where certain physical quantities such as density and curvature
become infinite. Singularities are associated with BHs, where the
curvature of spacetime becomes so extreme that it traps everything
including light within an event horizon. The concept of ``cosmic
censorship" was proposed by physicist Roger Penrose in the 1960s as
a way to protect the predictability and stability of the universe
\cite{6}. The Penrose cosmic censorship hypothesis suggests that
singularities resulting from the gravitational collapse of massive
objects are always hidden behind event horizons. In other words, the
extreme curvature and infinite density of a singularity are never
directly visible to observers outside the BH. There is no connection
between the interior (inside the EH) and the exterior (outside the
EH) regions of a BH as the physical laws hold in the exterior part.
Sakharov and Gliner \cite{7} suggested that one can avoid this
problem by a source of matter replaced with a de Sitter core at the
center of a BH. Bardeen \cite{8} found a BH by using a de Sitter
patch in the place of singularity, so called regular BH. Later many
non-singular BH models were proposed in the literature
\cite{9}-\cite{11}. Bronnikov \cite{12}, Dymnikova \cite{13} and
Hayward \cite{14} regular BHs are non-singular spherically symmetric
BHs that violate strong energy condition. Some axially symmetric and
rotatory regular BHs can also be found in the literature which
violate weak energy condition \cite{15}. Rincon and Santos \cite{16}
computed the GBF and quasinormal modes of non-singular BH solutions
and found that these modes are unstable.

The study of emission of Hawking radiations has gained the attention
of various researchers. Konoplya \cite{17} formulated the
quasinormal modes and potential barrier associated with the emission
of scalar field. The GBF and emission rate of BH produced in extra
dimensions were studied in \cite{18} which indicate that both
increase with extra dimensions. Grain et al \cite{19} computed GBF
of a BH for modified gravity and concluded that the effect of
Guass-Bonnet coupling constant shows different behavior in high and
low energy regions. Creek et al \cite{20} formulated an analytical
solution of rotating BH in the brane-scalar field for the GBF in the
high and low energy regions. They found that the numerical and
analytical results are comparable with each other. Harmark et al
\cite{21} computed the GBF for de Sitter and anti-de Sitter (AdS)
BHs in the presence of charge and cosmological constant.

Srinivasan and Padmanabhan \cite{22} discussed the GBF of
Schwarzschild-like spacetimes in the uniform electric field by the
complex path method. Jiang \cite{23} formulated the absorption
probability for the RN BH by using the tunneling method. Ngampitipan
and Boonserm \cite{24} found bounds on the absorption probability of
Reissner-Nordstr$\ddot{o}$m (RN) and Kerr-Newman BHs.  Kanti et al
\cite{25} computed the GBF of a scalar field in the higher
dimensions for Schwarzschild-de Sitter BH. Toshmatov et al \cite{26}
formulated the GBF of a regular charged BH and concluded that charge
decreases the absorption probability of Hawking radiations. The
emission rate of massive particles propagating from charged BH was
computed in \cite{27}. Ahmad and Saifullah \cite{28} formulated an
analytical expression of the GBF in the low energy regime and
developed a general expression of GBF for RN BH. Sharif and
Ama-Tul-Mughani \cite{29} found the GBF for a rotating BH surrounded
by quintessence energy and found that this energy enhances the GBF.

Panotopoulos and Rincon \cite{30} analyzed the GBF with minimally
coupled massless scalar field in various regimes and found that the
obtained results are comparable with previous work in those fields.
$\ddot{O}$vg$\ddot{u}$n and Jusufi \cite{31} computed the emission
rate of a massless scalar field and obtained that the behavior of
GBF is similar to Bekenstein's solution. Ahmad and Saifullah
\cite{32} examined the GBF for a non-minimally coupled scalar field
in RN-de Sitter regime and found that non-minimally coupled scalar
field decreases the emission rate. The absorption rate of BH for
non-minimally coupled massless scalar field with strings was studied
in \cite{33}. Ali et al \cite{34} studied the GBF of a spherically
symmetric charged regular de Sitter BH in higher dimensions. They
found that the non-minimal coupling reduces the GBF while the
non-linear charge increases the GBF. Several minimally and
non-minimally coupled BHs in de Sitter spacetime have been studied
with different considerations which give life span of the BHs
\cite{35}-\cite{37}. Recently, we have studied the GBF of a static
spherically symmetric BH with non-linear electrodynamics and found
that the non-linear charge parameter increases the absorption
probability of the BH \cite{38}.

Researchers have made progress in understanding the structure,
lifespan and information loss paradox associated with BHs. This
progress is achieved through the study of GBF in different regimes,
including de Sitter and AdS spacetimes. This exploration helps them
delve into the quantum gravity and quantum structure of BHs in
various regions. Born and Infeld \cite{39} used electrodynamics to
demonstrate that point-like charge possesses finite energy. Beato
and Garcia \cite{40} extended this concept by exploring the coupling
of electrodynamics with Einstein's theory and found non-singular BH
solutions. Black holes that are minimally or non-minimally coupled
with gravity in different regimes have attracted significant
attention from researchers. These investigations contribute to a
deeper understanding of the behavior of BHs. Motl and Neitzke
\cite{41} proposed that studying the GBF of Reissner-Nordstrom and
Kerr BHs can unveil new aspects of their quantum nature. This
exploration offers insights into the quantum characteristics of
these types of BHs. The study of GBF is crucial for understanding
the information loss paradox that arises from thermodynamics of BHs.
This paradox is linked to the study of Hawking radiation, which is
considered the carrier of information escaping from BHs. Researchers
are using various approaches, including electrodynamics and gravity
coupling, to advance our understanding of these enigmatic cosmic
objects.

In this paper, we compute the potential barrier and GBF for a non
accelerated BH with modified Maxwell electrodynamics in the AdS
regime. The paper is planned as follows. Section \textbf{2}
describes the solution of BH and formulates the effective potential
through radial equation. In section \textbf{3}, we discuss exact
solutions of the Regge-Wheeler equation at the event and
cosmological horizons. We match these solutions in an intermediate
region and formulate the expression for GBF in section \textbf{4}.
The summary of the main results is given in the last section.

\section{Non-Accelerated Black Hole with Modified Maxwell Electrodynamics}

The Einstein-Hilbert action of non-linear electrodynamics is defined
as \cite{42}
\begin{equation}\nonumber
I=\frac{1}{6\pi}\int_{M}{d^4x(R+\frac{6}{p^2}-4L)\sqrt{-g}},
\end{equation}
where $R$, $L$, $p$ and $g$ represent the Ricci scalar, Lagrangian
related to non-electrodynamics theory, AdS radius and determinant of
the line element, respectively. The Lagrangian depends on the
following two invariants
\begin{eqnarray}\nonumber
\quad S=\frac{F_{ab}F^{ab}}{2}, \quad P=\frac{F_{ab}(\ast
F)^{ab}}{2},
\end{eqnarray}
where $F_{ab}=\partial_a A_a-\partial_b A_a$ and $(\ast F)_{ab}=
\frac{\epsilon_{ab}F}{2}$ with vector potential $A_{a}$. The
Einstein-non linear electrodynamics equations are
\begin{eqnarray}\nonumber
\quad G_{ab}=8\pi T_{ab}, \quad d\ast E=0, \quad dF=0,
\end{eqnarray}
where $T_{ab}$ and $E_{ab}$ denote energy-momentum tensor and
non-linear function of $(\ast F_{ab},~F_{ab})$, respectively. They
are evaluated as
\begin{eqnarray}\nonumber
T_{ab}=\frac{1}{8\pi}\left(4F_{ac}F^{c}_{b}+2(PL_P-L)g_{ab}\right),\quad
E_{ab}=\frac{\partial L}{F_{ab}}=2\left(L_SF_{ab}+L_p\ast
F_{ab}\right),
\end{eqnarray}
where $L_S=\frac{\partial L}{\partial S}$ and $L_P=\frac{\partial
L}{\partial P}$. The Lagrangian for modified Maxwell theory is
expressed as \cite{43}
\begin{equation}\nonumber
L=\frac{1}{2}\left(S\cosh\gamma-\sqrt{S^2+P^2}\sinh\gamma\right),
\end{equation}
which is characterized by a dimensionless parameter $\gamma$ that
represents the modified Maxwell BH solution.

The respective line element of non-accelerated BH with modified
Maxwell electrodynamics in AdS regime is \cite{42}
\begin{equation}\label{1}
ds^2=-\frac{f(r)dt^{2}}{\alpha^2}+\frac{dr^2}{f(r)}+r^2(d\theta^{2}+\sin^2\theta
\frac{d\phi^2}{K^2}),
\end{equation}
where $\alpha$ is the real constant and
\begin{equation}\label{2}
f(r)=1-\frac{2m}{r}+\frac{z^2}{r^2}+\frac{r^2}{p^2},
\end{equation}
where $z^2=e^{-\gamma}(q_{m}^{2}+q_{e}^{2})$. Here $m$, $r$, $K$,
$q_m$ and $q_e$ are mass, radius, conical deficits, magnetic and
electric charge of the BH, respectively. The given BH solution is
singular at $r=p=0$. A static spherically symmetric BH is obtained
for $\alpha=1$ as well as reduces to Schwarzschild BH for $z=0$ and
$p\rightarrow\infty$. The EHs of the given BH are obtained by taking
\begin{equation}\label{3}
1-\frac{2m}{r}+\frac{z^2}{r^2}+\frac{r^2}{p^2}=0.
\end{equation}

\subsection{The Effective Potential}

In this section, we compute the potential required to cross the
gravitational barrier. We examine the propagation of massless scalar
field by using the Klein-Gordon equation as
\begin{equation}\label{4}
\partial_{a}[\sqrt{-g}g^{ab}\partial_{b}\Psi]=0,
\end{equation}
where $g$ is the determinant of the line element. We assume that
particles and gravity are minimally coupled during the propagation
of scalar field. We separate Eq.(\ref{4}) into two wave equations by
using the separation of variables method
\begin{equation}\nonumber
\Psi=e^{iwt}R_{wlm}(r)A^{l}_{m}(\theta,\phi),
\end{equation}
where $A^{l}_{m}(\theta,\phi)$ is an angular function. The function
$R_{wlm}(r)$ is the solution of the following radial equation
\cite{44}
\begin{equation}\label{5}
\left(r^2f\frac{d}{dr}R_{wlm}\right)_{,r}+\left(\frac
{r^2\alpha^2w^2}{f}-\lambda_{l}\right)R_{wlm}=0,
\end{equation}
where $\lambda_{l}=l^2+l$ determines the relation between angular
and radial equations \cite{45}. We can solve Eq.(\ref{5}) to
formulate the GBF of the massless scalar field. Firstly, we find the
effective potential which affects the GBF of the BH. Therefore, we
use the new radial equation
\begin{equation}\label{6}
R_{wlm}(r)=\frac{Y_{wlm}(r)}{r},
\end{equation}
and tortoise coordinate as
\begin{eqnarray}\nonumber
\quad\frac{du_{\star}}{dr}=\frac{1}{f},
\quad\frac{d}{du_{\star}}=f\frac{d}{dr},\quad\frac{d^2}{du_{\star}^2}=f
\left(\frac{d^2}{dr^2}+\frac{df}{dr}\frac{d}{dr}\right).
\end{eqnarray}
We note that $r$ approaches to $r_h$, tortoise coordinate approaches
to $-\infty$ and $u_\star\rightarrow \infty$ as $r \rightarrow
\infty$. The radial equation works for the whole real line due to
the tortoise coordinate, hence the radial equation is written as
\begin{equation}\label{7}
(\frac{d^{2}}{du_\star^{2}}-V_{ef})Y_{wlm}=0,
\end{equation}
where the effective potential
\begin{equation}\nonumber
V_{ef}=f\left(\frac{1}{r}\frac{df}{dr}-w^2\alpha^2+\frac{\lambda_{l}}{r^2}\right),
\end{equation}
which vanishes at $f=0$. The explicit form of the effective
potential turns out to be
\begin{eqnarray}\nonumber
&&V_{ef}=\frac{1}{r^{6}p^{4}}\bigg(r^4+p^2r^2-2mp^2r+p^2z^2\bigg)
\bigg((2-\alpha^2p^2w^2)r^4\\\nonumber
&&+(l^2p^2+lp^2)r^2+2mp^2r+2p^2z^2\bigg).
\end{eqnarray}
We analyze the graphical behavior of
effective potential against $\frac{r}{r_h}$ corresponding to
different parameters $(q_{m},~q_{e},~m,~l,~\alpha,~p,~w,~\gamma)$.
Figure \textbf{1} shows that the effective potential has a direct
relation with $q_{m}$ and $q_{e}$ which shows the decrease in
absorption probability of the BH. Figure \textbf{2} represents that
the gravitational barrier decreases with the increase in mass and
decrease in angular momentum which shows the enhancement in
evaporation rate. In Figure \textbf{3}, the graph shows that
gravitational barrier has an inverse relation with $\alpha$ and AdS
radius which shows the increasing behavior of the GBF. The graphical
behavior of wave frequency and the parameter characterizing the
modification of the Maxwell solution is given in Figure \textbf{4}.
The graph in the left panel shows that there is an inverse relation
of effective potential for different values of wave frequency
against $\frac{r}{r_h}$ which indicates that frequency parameter
enhances the GBF. The right plot describes that the height of the
graph decreases with an increase in $\gamma$. This behavior shows
that the modified Maxwell theory enhances the absorption probability
of the BH and consequently maximizes the evaporation process.
\begin{figure}
\epsfig{file=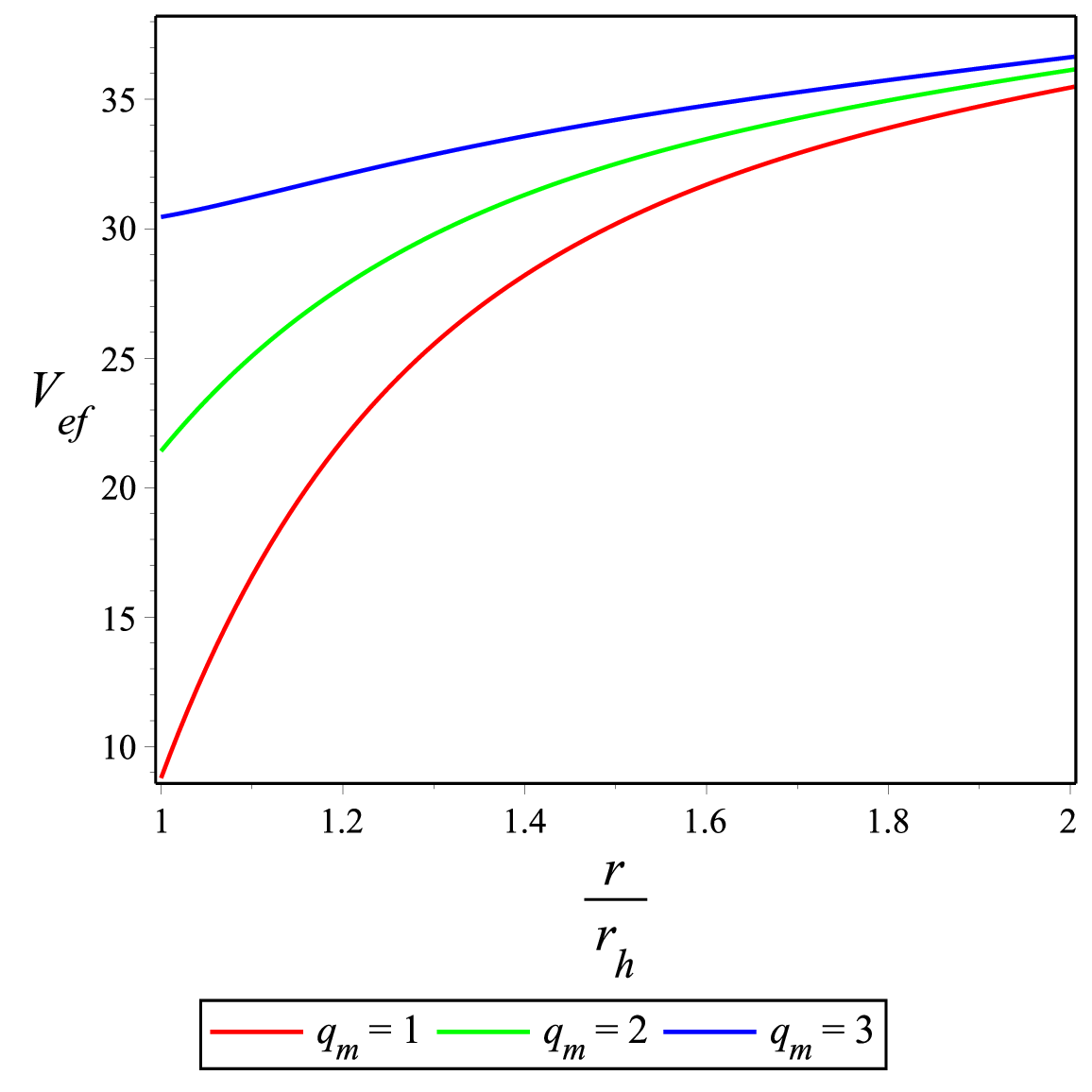,width=0.5\linewidth}
\epsfig{file=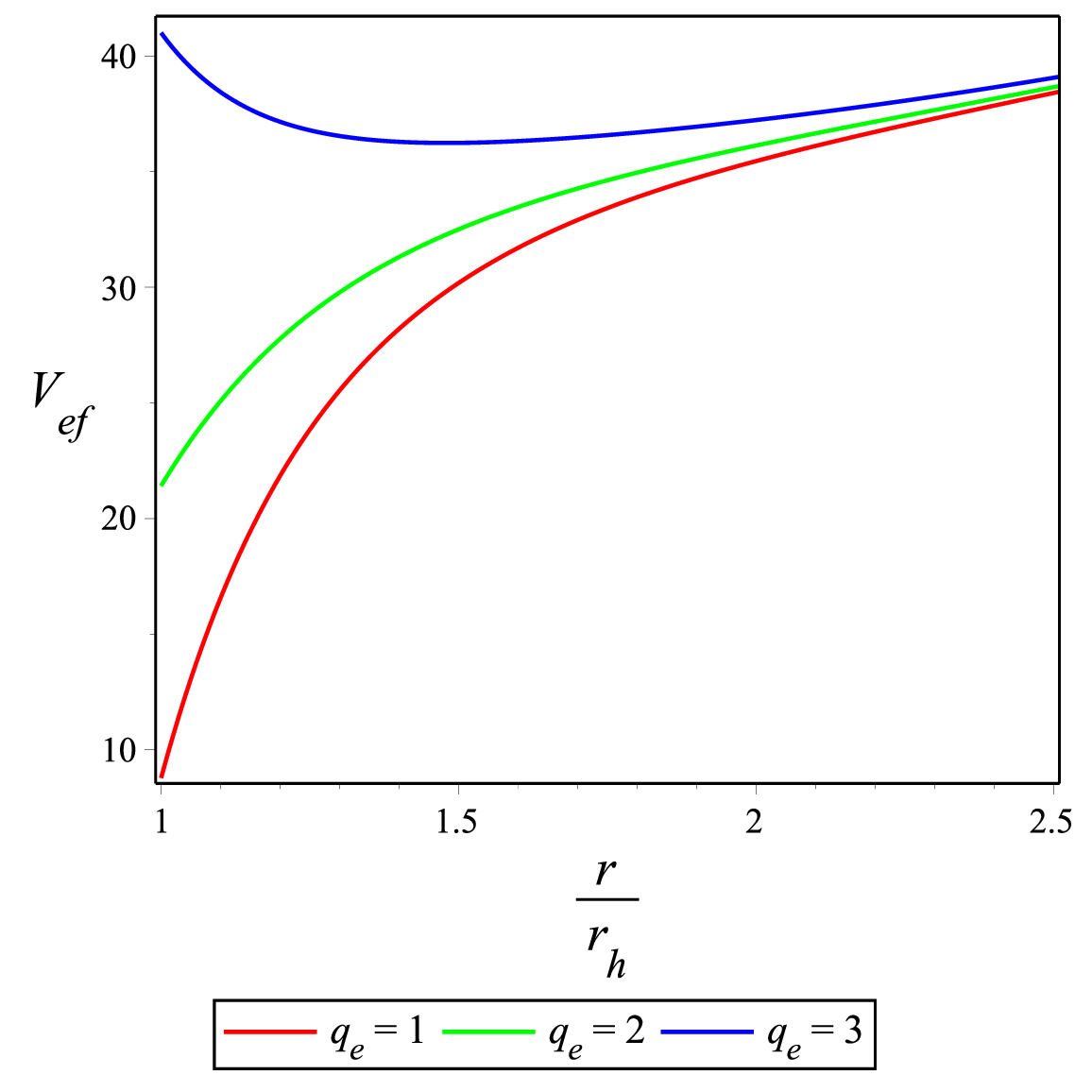,width=0.5\linewidth}\caption{Effective potential
versus $\frac{r}{r_{h}}$ with $q_{e}=1$ (left) and $q_{m}=1$ (right)
for $m=5$, $l=3$, $p=1$, $\alpha=1$, $\gamma=2$ and $w=1$.}
\end{figure}
\begin{figure}
\epsfig{file=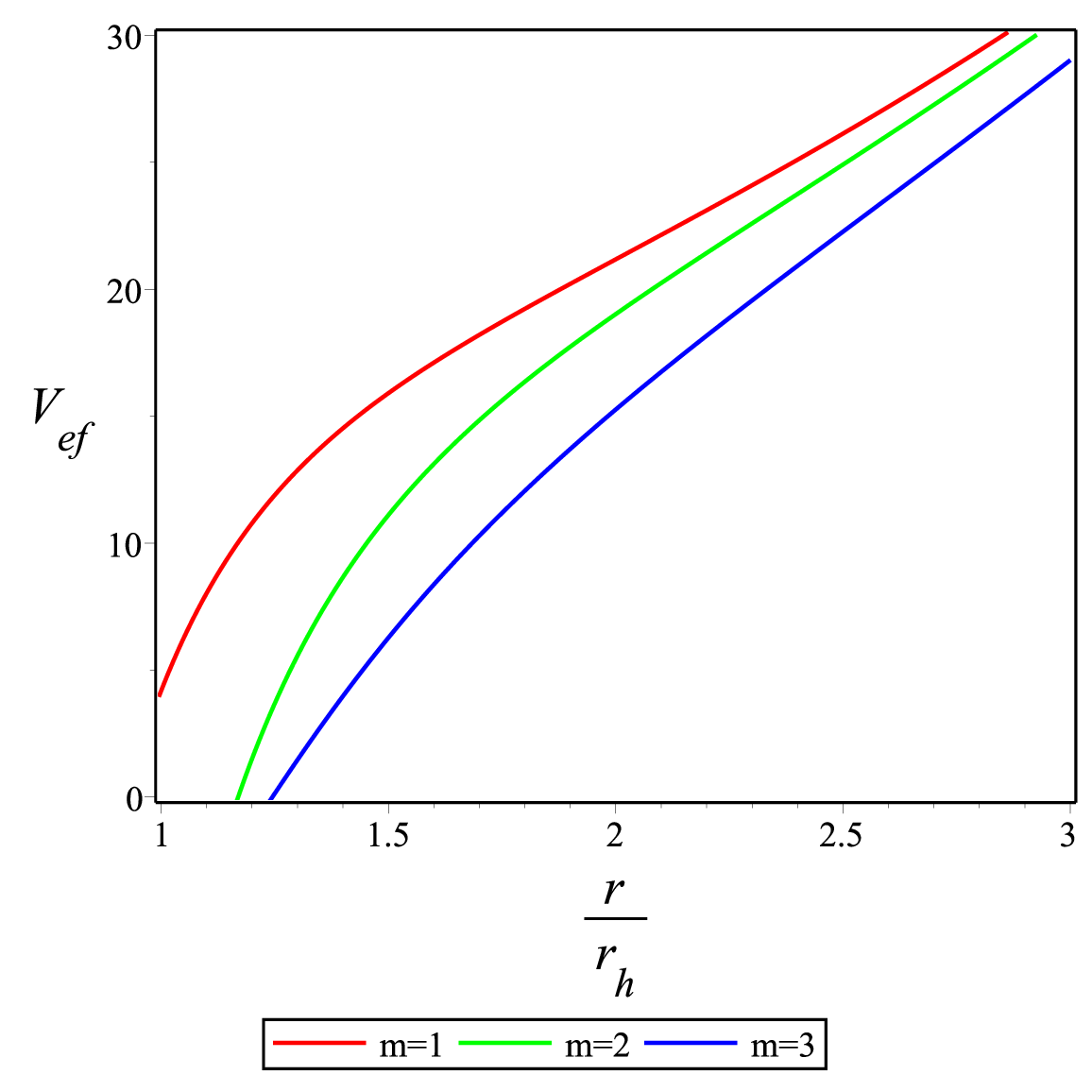,width=0.5\linewidth}
\epsfig{file=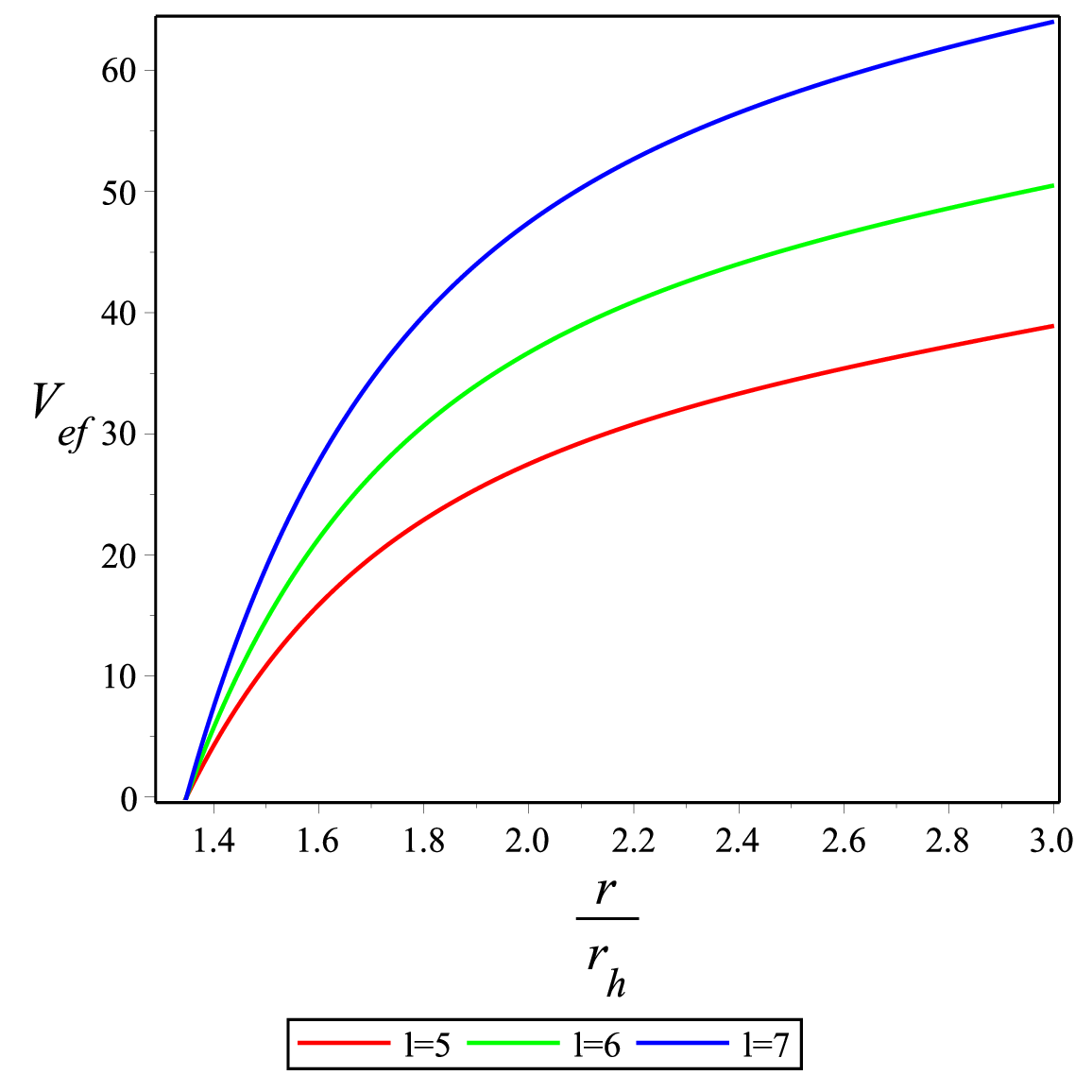,width=0.5\linewidth}\caption{Effective potential
versus $\frac{r}{r_{h}}$ with $l=3$ (left) and $m=5$ (right) for
$q_{e}=1$, $q_m=1$, $p=1$, $\alpha=0.1$, $\gamma=2$ and $w=1$.}
\end{figure}
\begin{figure}
\epsfig{file=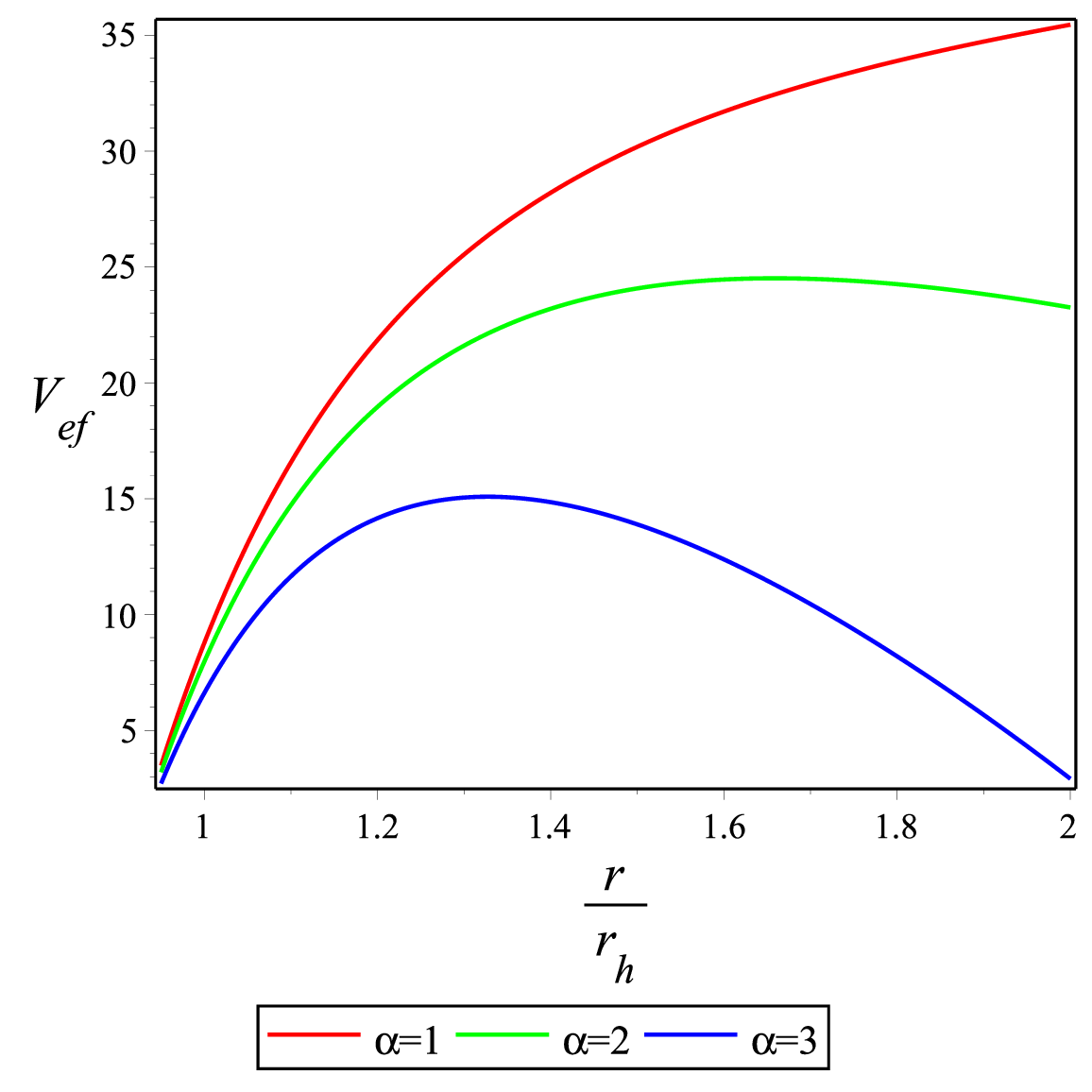,width=0.5\linewidth}
\epsfig{file=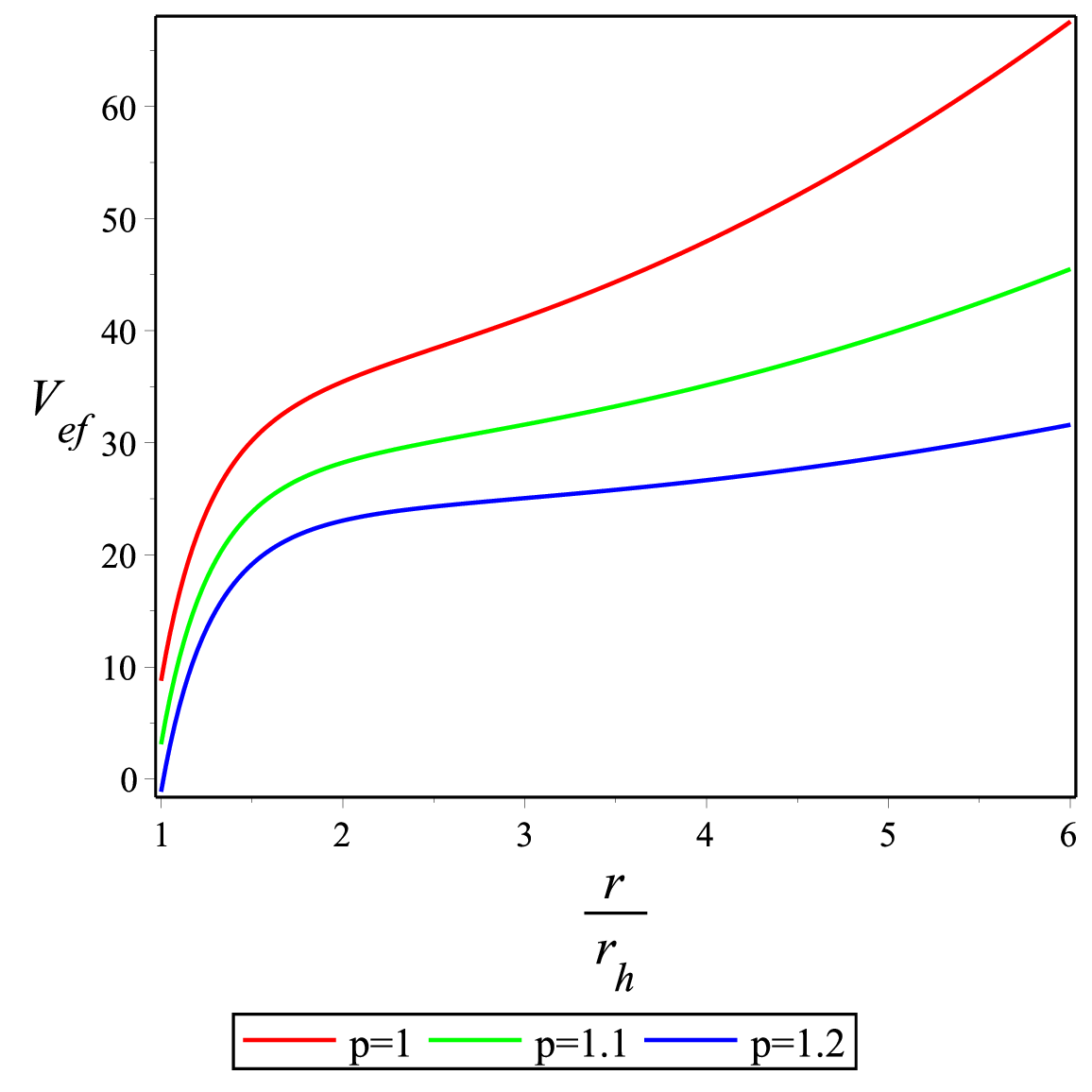,width=0.5\linewidth}\caption{Effective potential
versus $\frac{r}{r_{h}}$ with $p=1$ (left) and $\alpha=1$ (right)
for $m=5$, $l=3$, $q_e=1$, $q_m=1$, $\gamma=2$ and $w=1$.}
\end{figure}
\begin{figure}
\epsfig{file=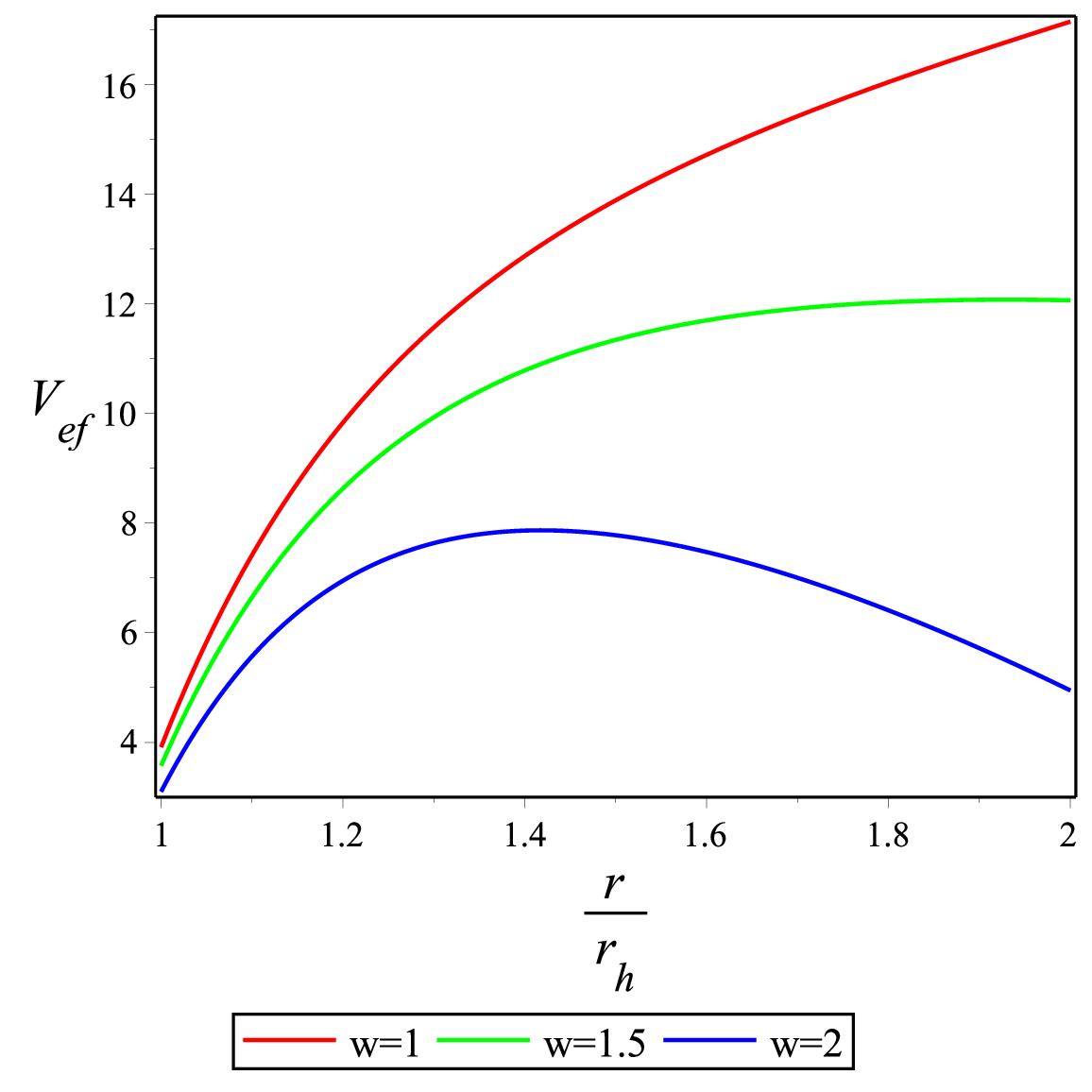,width=0.5\linewidth}
\epsfig{file=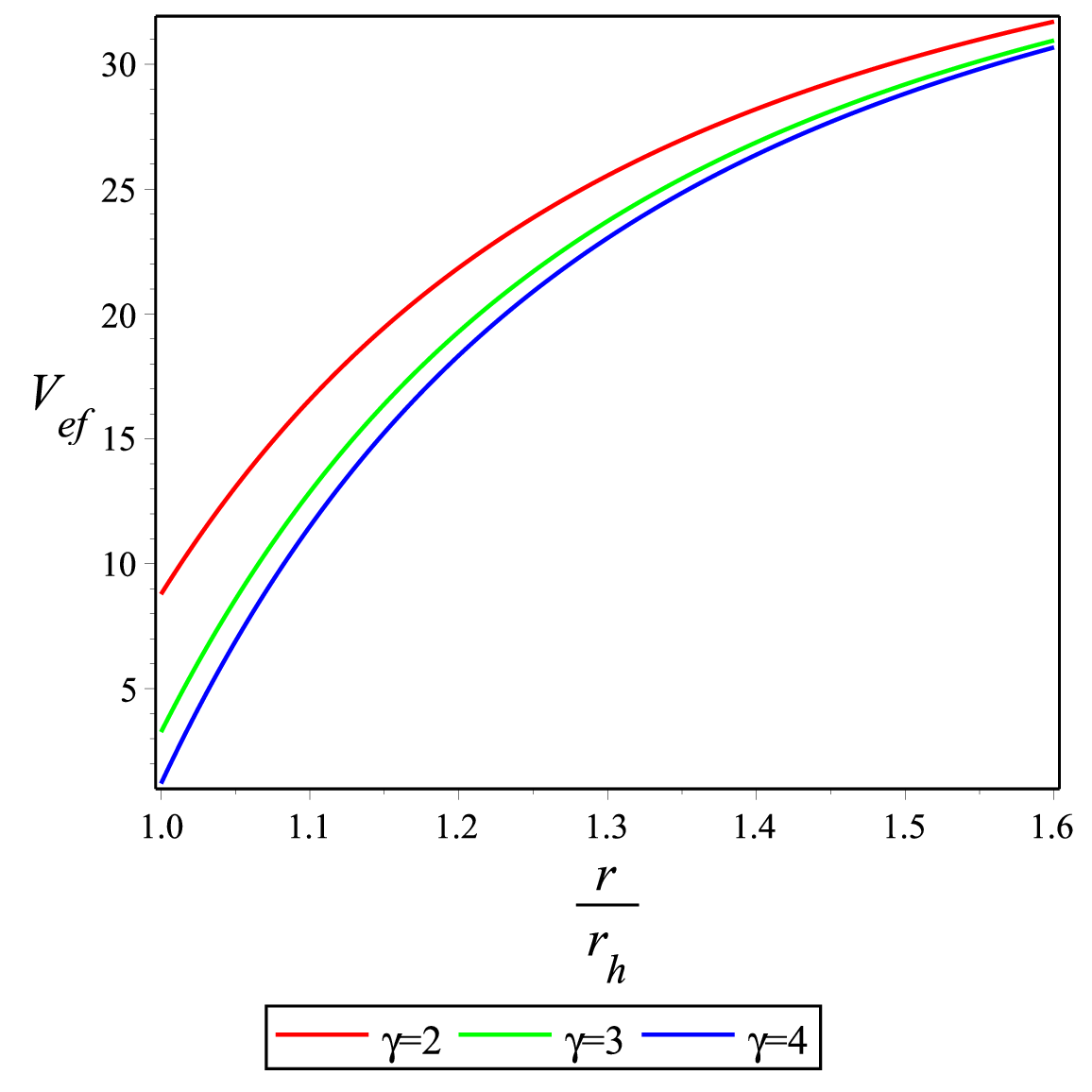,width=0.5\linewidth}\caption{Effective potential
versus $\frac{r}{r_{h}}$ with $\gamma=2$ (left) and $w=1$ (right)
for $m=5$, $l=3$, $p=1$, $\alpha=1$, $q_m=1$ and $q_e=1$.}
\end{figure}

\section{Analytic Solutions and Greybody Factor}

This section uses transformations to obtain analytical solutions of
Eq.(\ref{5}) at the event as well as cosmological horizon. Further,
we apply a semi-classical method to match these solutions in the
intermediate region. These transformations give equations in the
form of hypergeometric (HG) function whose solutions are used to
obtain an expression of GBF. For the region near to the EH $(r\sim
r_{h})$, we use the first transformation as
\begin{equation}\label{8}
\Psi=\frac{f}{1+\frac{r^2}{p^2}}=\frac{1-\frac{2m}{r}+\frac{z^2}{r^2}+\frac{r^2}
{p^2}}{1+\frac{r^2}{p^2}},
\end{equation}
which satisfies the relation
\begin{equation}\nonumber
\frac{d\Psi}{dr}=\frac{(1-\Psi)B} {r(p^2+r^{2})},
\end{equation}
where
\begin{equation}\label{9}
B=\frac{2p^2(3mr^3-2r^2z^2+(mr-z^2)p^2)}{2mr-z^2}.
\end{equation}
Using Eq.(\ref{9}) in the radial equation, we have
\begin{equation}\label{10}
\Psi(1-\Psi)\frac{d^{2}R_{wlm}}{d\Psi^{2}}+(C-D\Psi)\frac{dR_{wlm}}
{d\Psi}+\frac{1}{(1-\Psi)B^2}(\frac{\zeta_{h}}{\Psi}-\xi_{h})R_{wlm}=0,
\end{equation}
where
\begin{eqnarray}\nonumber
C=\frac{p^2(rf(1-\Psi)B)'}{(1-\Psi)B^{2}},\quad
D=-\frac{2r^{2}}{B},~\zeta_{h}=p^4w^{2}r^{2}\alpha^2,
~\xi_{h}=p^2(l^2+l)(p^2+r^{2}).
\end{eqnarray}
We use the field redefinition
$(R_{wlm}(\Psi)=\Psi^{\alpha_{1}}(1-\Psi)^{\beta_{1}}W_{wlm}(\Psi)),$
which reduces Eq.(\ref{10}) to HG differential equation as
\begin{eqnarray}\nonumber
&&\Psi(1-\Psi)\frac{d^2W_{wlm}}{d\Psi^2}+[2\alpha_{1}+C-(2\alpha_1+2
\beta_1+D)\Psi]\frac{dW_{wlm}}{d\Psi}+\bigg[\frac{1}{\Psi}(\alpha_{1}
^{2}-\alpha_1\\\nonumber&&+C\alpha_1+\frac{\zeta_h}{B^2})+\frac{1}{1-\Psi}
(\beta_{1}^{2}-\beta_1-C\beta_1+D\beta_1+\frac{\zeta_h}{B^2}-\frac{\xi_h}
{B^2})-(\beta_1+\alpha_1)A\\\nonumber&&-\alpha_{1}^{2}-2\alpha_1\beta_1+\alpha_1-
\beta_{1}^{2}+\beta_1\bigg]W_{wlm}=0.
\end{eqnarray}
The power coefficients $\alpha_{1}$ and $\beta_{1}$ can be found by
solving the following equations
\begin{eqnarray}\label{11}
\alpha_{1}^{2}-\alpha_{1}(1-C)+\frac{\zeta_{h}}{B^{2}}=0,\quad\beta_{1}^{2}
-\beta_1(1+C-D)+\frac{\zeta_{h}}{B^{2}}-\frac{\xi_{h}}{B^2}=0.
\end{eqnarray}

The corresponding radial equation (\ref{5}) together with
Eq.(\ref{11}) leads to
\begin{equation}\label{12}
\Psi(1-\Psi)\frac{d^2W_{wlm}}{d\Psi^2}+(x_1-(t_{1}+y_{1}+1)
\Psi)\frac{dW_{wlm}}{d\Psi}-t_{1}y_{1}W_{wlm}=0,
\end{equation}
where $t_{1}=\alpha_{1}+\beta_{1}$,
$y_{1}=\alpha_{1}+\beta_{1}+D-1$, $x_1=2\alpha_{1}+C$. The general
solution of Eq.(\ref{12}) in terms of HG function is
\begin{eqnarray}\nonumber
(R_{wlm})_{nh}(\Psi)&=&T_{1}\Psi^{\alpha_{1}}(1-\Psi)^{\beta_{1}}
F(t_{1},y_{1},x_1;\Psi)+T_{2}\Psi^{-\alpha_{1}}(1-\Psi)
^{\beta_{1}}\\\nonumber&\times&F(1-x_1+t_{1},1-x_1+y_{1},2-x_1
;\Psi),
\end{eqnarray}
where $T_1$ and $T_2$ are integration constants and
\begin{eqnarray}\nonumber
\alpha_{1}^{\pm}&=&\frac{1}{2}\left[(1-C)\pm\sqrt{(1-C)^{2}-4\frac{\zeta_h}
{B^2}}\right],\\\nonumber\beta_{1}^{\pm}&=&\frac{1}{2}\left[(1+C-D)\pm
\sqrt{(1-D+C)^2+4(\frac{\xi_h}{B^2}\frac{\zeta_h}{B^2})}\right].
\end{eqnarray}
Applying the boundary conditions by which no outgoing mode is
observed at $r\sim r_{h}$, therefore, we may take either $T_1=0$, or
$T_2=0$ depending upon the choice of signature of $\alpha_1$. Here,
we take $\alpha_{1}=\alpha_{1}^{-}$ with $T_2=0$. Hence, the final
form of the corresponding solution is given as
\begin{equation}\label{13}
(R_{wlm})_{nh}(\Psi)=T_{1}\Psi^{\alpha_{1}}(1-\Psi)^{\beta_{1}}
F(t_{1},y_{1},x_1;\Psi).
\end{equation}

Now, we obtain solution of the radial equation at the cosmological
horizon. Applying the same procedure, we have
\begin{equation}\label{14}
\Omega(r)=\frac{f}{r^{2}}=\frac{1}{r^2}+\frac{1}{p^2},
\end{equation}
which gives
\begin{equation}\label{15}
\frac{d\Omega}{dr} =\frac{(1-\Omega)F}{r},
\end{equation}
where $F(r)=\frac{-2p^2}{r^2+p^2}$. Using the transformation
$R_{wlm}(\Omega)=\Omega^{\alpha_{2}}(1-\Omega)^{\beta_{2}}W_{wlm}(\Omega)$,
in radial equation, it follows that
\begin{eqnarray}\nonumber
&&\Omega(1-\Omega)\frac{d^2W_{wlm}}{d\Omega^2}+(2\alpha_{2}+C_\star
-(2\alpha_2+2\beta_2+D_\star)\Omega)\frac{dW_{wlm}}{d\Omega}\\\nonumber&&
+\bigg[(\alpha_{2}^{2}-\alpha_2+C_{\star}\alpha_2+\frac{\zeta_\star}{F^2})\frac{1}
{\Omega}+(\beta_{2}^{2}-\beta_2-C_{\star}\beta_2+D_{\star}\beta_2+\frac{\zeta
_\star}{F^2}-\frac{\xi_\star}{F^2})\\\label{16}&&\times\frac{1}{1-\Omega}
-\alpha_{2}^{2}-2\alpha_2\beta_2-D_{\star}(\alpha_2-\beta_2+\alpha_2+\beta_2)\bigg]W_{wlm}=0.
\end{eqnarray}
The power coefficients $\alpha_2$ and $\beta_2$ can be obtained by
solving the following equations
\begin{eqnarray}\nonumber
\alpha_{2}^{2}-\alpha_{2}(1-C_\star)+\frac{\zeta_{\star}}{F^{2}}=0,\quad\beta_{2}^{2}
-\beta_{2}(1-D_{\star}+C_{\star})+\frac{\zeta_{\star}}{F^{2}}-\frac{\xi_{\star}}
{F^2}=0.
\end{eqnarray}
At cosmological horizon, Eq.(\ref{16}) reduces to the HG
differential equation as
\begin{equation}\label{17}
\Omega(1-\Omega)\frac{d^2W_{wlm}}{d\Omega^2}+(x_{2}-(t_{2}+y_{2}+1)\Omega)
\frac{dW_{wlm}}{d\Omega}-t_{2}y_{2}W_{wlm}=0,
\end{equation}
where $t_{2}=\alpha_{2}+\beta_{2},
y_{2}=\alpha_{2}+\beta_{2}+D_{\star}-1,
x_{2}=2\alpha_{2}+C_{\star}$. The general solution of HG equation is
\begin{eqnarray}\nonumber
(R_{wlm})_{fh}(\Omega)&=&Y_{1}\Omega^{\alpha_{2}}(1-\Omega)^{\beta_{2}}F(t_{2},y_{2}
,x_{2};\Omega)+Y_{2}\Omega^{-\alpha_{2}}(1-\Omega)^{\beta_{2}}\\\label{18}
&\times&F(1+t_{2}-x_{2},1+y_{2}-x_{2},2-x_{2};\Omega),
\end{eqnarray}
where $Y_1$ and $Y_2$ are integration constants.

\section{Matching Regime}

We must match $(R_{wlm})_{nh}(\Psi)$ and $(R_{wlm})_{fh}(\Omega)$ in
the intermediate region of radial coordinate to get the analytical
solution for the whole range of $r$. Therefore, we first stretch the
EH towards the intermediate region by replacing $\Psi$ by $1-\Psi$
in Eq.(\ref{13}) as
\begin{eqnarray}\nonumber
(R_{wlm})_{nh}(\Psi)&=&T_{1}\Psi^{\alpha_{1}}(1-\Psi)^{\beta_{1}}\bigg[\frac
{\Gamma(-t_{1}-y_{1}+x_1)\Gamma(x_{1})}{\Gamma(x_1-t_{1})\Gamma(x_{1}-y_{1}
)}F(t_{1},y_{1},x_{1};1-\Psi)\\\nonumber&+&(1-\Psi)^{-t_1-y_1+x_1}\frac{
\Gamma(x_1)\Gamma(t_1+y_1-x_1)}{\Gamma(y_1)\Gamma(t_1)}\\\nonumber&\times&
F(x_1-t_1,x_1-y_1,1-t_1-y_1+x_1;1-\Psi)\bigg].
\end{eqnarray}
Using Eqs.(\ref{3}) in (\ref{8}), we obtain
\begin{equation}\nonumber
1-\Psi=\frac{p^2(2mr-z^2)}{p^2+r^2}.
\end{equation}
The stretched EH for the limiting value ($r\gg r_h$ and
$\Psi\rightarrow{1}$) takes the form
\begin{eqnarray}\nonumber
(1-\Psi)^{\beta_1}\simeq
\left(\frac{z_{\star}^{2}p^2+\frac{1}{r_{h}^{2}}}{p^2+r^2}\right)^{\beta_1}
\left(\frac{r}{r_h}\right)^{\beta_1}\Rightarrow(1-\Psi)^{\beta_1}\simeq
\left(\frac{z_{\star}^{2}p^2+\frac{1}{r_{h}^{2}}}{p^2+r^2}\right)^{-l}\left(
\frac{r}{r_h}\right)^{-l},
\end{eqnarray}
and
\begin{eqnarray}\nonumber
(1-\Psi)^{\beta_1+x_1-t_1-y_1}&\simeq&\left(\frac{z_{\star}^{2}p^2+\frac{1}
{r_{h}^{2}}}{p^2+r^2}\right)^{-\beta_1+B-A+1}\left(\frac{r}{r_h}\right)^{-
\beta_1+C-D+1}\\\nonumber&\simeq&
\left(\frac{z_{\star}^{2}p^2+\frac{1}{r_{h}^{2}}}{p^2+r^2}\right)^{l+1}\left(
\frac{r}{r_h}\right)^{l+1},
\end{eqnarray}
where $z_{\star}^{2}=\frac{z^2}{r_{h}^{2}}$. Here we assume that the
values of electric and magnetic charges are small. This assumption
makes our results valid in the low energy region. Both parts of near
horizon BH solution in the intermediate region can be written as
\begin{eqnarray}\nonumber
(1-\Psi)^{\beta_1}&\simeq&
\left(\frac{z_{\star}^{2}p^2+\frac{1}{r_{h}^{2}}}{p^2+r^2}\right)^{-l}\left(\frac{r}{r_h}\right)^{-l},
\\\nonumber
(1-\Psi)^{\beta_1+x_1-t_1-y_1}&\simeq&
\left(\frac{z_{\star}^{2}p^2+\frac{1}{r_{h}^{2}}}{p^2+r^2}\right)^{1+l}\left(\frac{r}{r_h}\right)^{1+l}.
\end{eqnarray}
Finally, the solution on the EH is
\begin{equation}\label{19}
(R_{wlm})_{nh}(\Psi)=T'_{1}(\frac{r}{r_h})^{-l}+T'_{2}(\frac{r}{r_h})^{l+1},
\end{equation}
with
\begin{eqnarray}\nonumber
T'_1&=&T_1\left(\frac{z_{\star}^{2}p^2+\frac{1}{r_{h}^{2}}}{p^2+r^2}\right)^
{-l}\frac{\Gamma(-t_{1}-y_{1}+x_1)\Gamma(x_1)}{\Gamma(x_1-t_{1})
\Gamma(x_1-y_{1})},\\\nonumber
T'_2&=&T_2\left(\frac{z_{\star}^{2}p^2+\frac{1}{r_{h}^{2}}}{p^2+r^2}\right)^
{l+1}\frac{\Gamma(x_1)\Gamma(t_1+y_1-x_1)}{\Gamma(y_1)\Gamma(t_1)}.
\end{eqnarray}

Now, we replace the argument $\Omega$ by $1-\Omega$ in Eq.(\ref{18})
for shifting the cosmological horizon to the intermediate region and
obtain
\begin{eqnarray}\nonumber
&&(R_{wlm})_{fh}(\Omega)=Y_1\Omega^{\alpha_{2}}(1-\Omega)^{\beta_{2}}
\bigg[\frac{\Gamma(-t_{2}-y_{1}+x_1)\Gamma(x_{2})}{\Gamma(x_{2}-t_{2})
\Gamma(x_{2}-y_{2})}F(t_{2},y_{2},x_{2};1-\Omega)\\\nonumber&&+F(x_2-t_2
,x_2-y_2,1-t_2-y_2+x_2;1-\Omega)\frac{\Gamma(x_2)\Gamma(t_2+y_2-x_2)}
{\Gamma(y_2)\Gamma(t_2)}\\\nonumber&&\times(1-\Omega)^{-t_2-y_2+x_2}
\bigg]+Y_{2}\Psi^{-\alpha_{2}}(1-\Omega)^{\beta_{2}}\bigg[\frac{\Gamma
(-t_{2}-y_{2}+x_{2})\Gamma(2-x_{2})}{\Gamma(1-t_{2})\Gamma(1-y_{2})}
\\\nonumber&&\times F(-x_2+t_2+1,y_2-x_2+1,2-x_2;1-\Omega)+(1-\Omega)
^{-t_2-y_2+x_2}\\\label{20}&&\times\frac{\Gamma(2-x_2)\Gamma(t_2+y_2-x_2)}
{\Gamma(1-t_2)\Gamma(1-y_2)}F(1-t_2,1-y_2,1-t_2-y_2+x_1;1-\Omega)\bigg].
\end{eqnarray}
By setting $\Omega(r_f)\rightarrow 0$, Eq.(\ref{14}) becomes
\begin{equation}\label{21}
1-\Omega=\frac{r}{r_{fh}}\left(\frac{1}{rr_{fh}}-\frac{r_f}{r^3}+\frac{r_{fh}}{r}\right),
\end{equation}
which can be written as
\begin{eqnarray}\nonumber
(1-\Omega)^{\beta_2}\simeq\left(\frac{r}{r_{fh}}\right)^{-l}\left(\frac{1}{rr_{fh}}
-\frac{r_{fh}}{r^3}+\frac{r_{fh}}{r}\right)^{-l},
\end{eqnarray}
and
\begin{eqnarray}\nonumber
(1-\Omega)^{\beta_2-t_2-y_2 +x_2
}\simeq\left(\frac{r}{r_{fh}}\right)^{l+1}\left(\frac{1}{rr_{fh}}-\frac{r_{fh}}{r^3}
+\frac{r_{fh}}{r}\right)^{l+1}.
\end{eqnarray}
The corresponding Eq.(\ref{20}) turns out to be
\begin{equation}\label{22}
R_{fh}=(S'_1Y_1+S'_2Y_2)\left(\frac{r}{r_{fh}}\right)^{-l}+(S'_3Y_1+S'_4Y_2)\left(\frac{r}
{r_{fh}}\right)^{l+1},
\end{equation}
where
\begin{eqnarray}\nonumber
&&S'_1=\frac{\Gamma(x_2)\Gamma(x_2-t_2-y_2)}{\Gamma(x_2-t_2)\Gamma(x_2-y_2)}
\left(\frac{1}{rr_{fh}}-\frac{r_{fh}}{r^3}+\frac{r_{fh}}{r}\right)^{-l},\\\nonumber
&&S'_2=\frac{\Gamma(2-x_2)\Gamma(x_2-t_2-y_2)}{\Gamma(1-t_2)\Gamma(1-y_2)}
\left(\frac{1}{rr_{fh}}-\frac{r_{fh}}{r^3}+\frac{r_{fh}}{r}\right)^{-l},\\\nonumber
&&S'_3=\frac{\Gamma(x_2)\Gamma(-x_2+t_2+y_2)}{\Gamma(t_2)\Gamma(y_2)}
\left(\frac{1}{rr_{fh}}-\frac{r_{fh}}{r^3}+\frac{r_{fh}}{r}\right)^{l+1},
\\\nonumber&&S'_4=\frac{\Gamma(2-x_2)\Gamma(-x_2+t_2+y_2)}{\Gamma(1-t_2)
\Gamma(1-y_2)}\left(\frac{1}{rr_{fh}}-\frac{r_{fh}}{r^3}+\frac{r_f}{r}\right)^{l+1}.
\end{eqnarray}

Comparing the coefficients of solutions (\ref{19}) and (\ref{22}),
we obtain
\begin{equation}\nonumber
T'_1=S'_1Y_1+S'_2Y_2,~~T'_2=S'_3Y_1+S'_4Y_2,
\end{equation}
where
\begin{eqnarray}\label{23}
Y_1=\frac{T'_1S'_4-T'_2S'_2}{S'_1S'_4-S'_2S'_3},\quad
Y_2=\frac{T'_1S'_3-T'_2S'_1}{S'_2S'_3-S'_1S'_4}.
\end{eqnarray}\\
Here $Y_1$ and $Y_2$ denote the ingoing and out-coming waves.
Consequently, using Eq.(\ref{23}) and
\begin{equation}\label{24}
|G_{w,l,m}|^{2}=1-\left|\frac{Y_2}{Y_1}\right|^{2},
\end{equation}
we have final expression for the GBF as \cite{16}
\begin{equation}\label{25}
|G_{w,l,m}|^{2}=1-\left|\frac{T'_1S'_3-T'_2S'_1}
{T'_1S'_4-T'_2S'_2}\right|^{2}.
\end{equation}
Finally, we obtain an expression of GBF for the non accelerated BH
with modified Maxwell electrodynamics. The Hawking radiations
passing through the cosmological horizon face the gravitational
barrier. Therefore, some of the radiations transmit towards the EH
while some reflect back to the cosmological horizon which is due to
the relation between frequency and potential. The wave can only
cross the barrier when the frequency of the wave is higher than the
threshold frequency (the minimum frequency below which a wave cannot
cross the barrier).
\begin{figure}
\epsfig{file=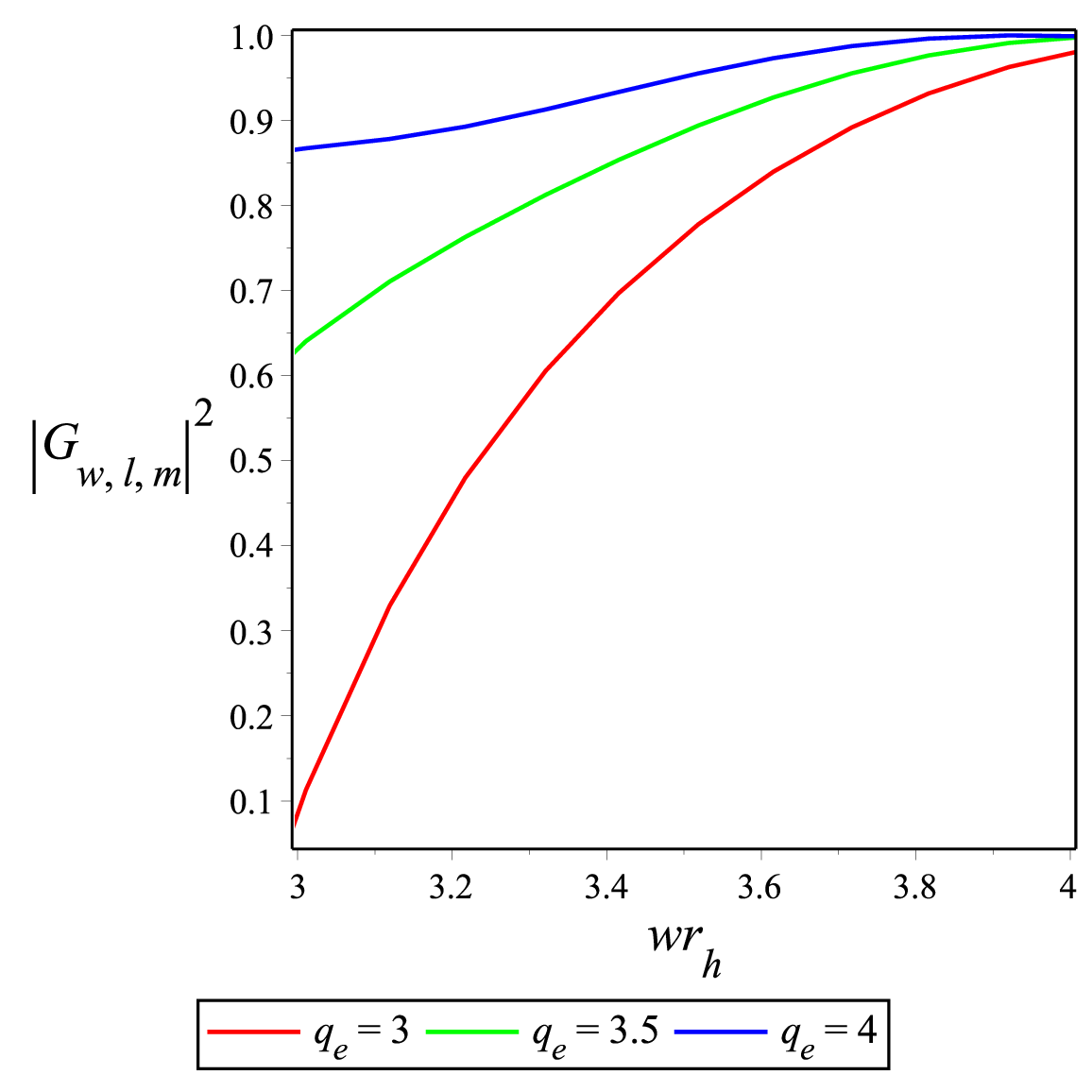,width=0.5\linewidth}
\epsfig{file=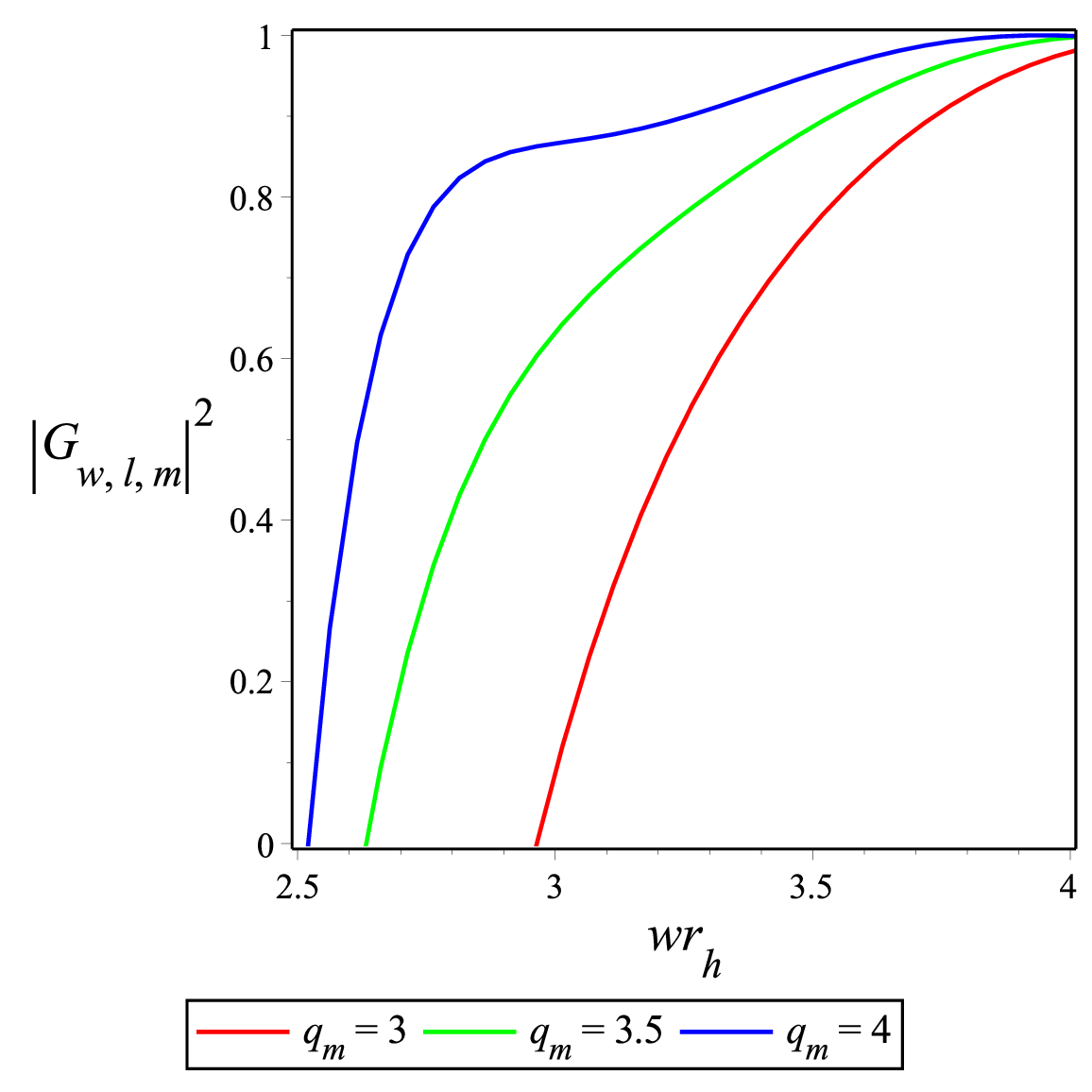,width=0.5\linewidth}\caption{GBF versus $wr_{h}$
with $q_m=3$ (left) and $q_e=3$ (right) for $m=5$, $l=5$, $p=1$,
$\alpha=1$, $\gamma=2$ and $r=1$.}
\end{figure}
\begin{figure}
\epsfig{file=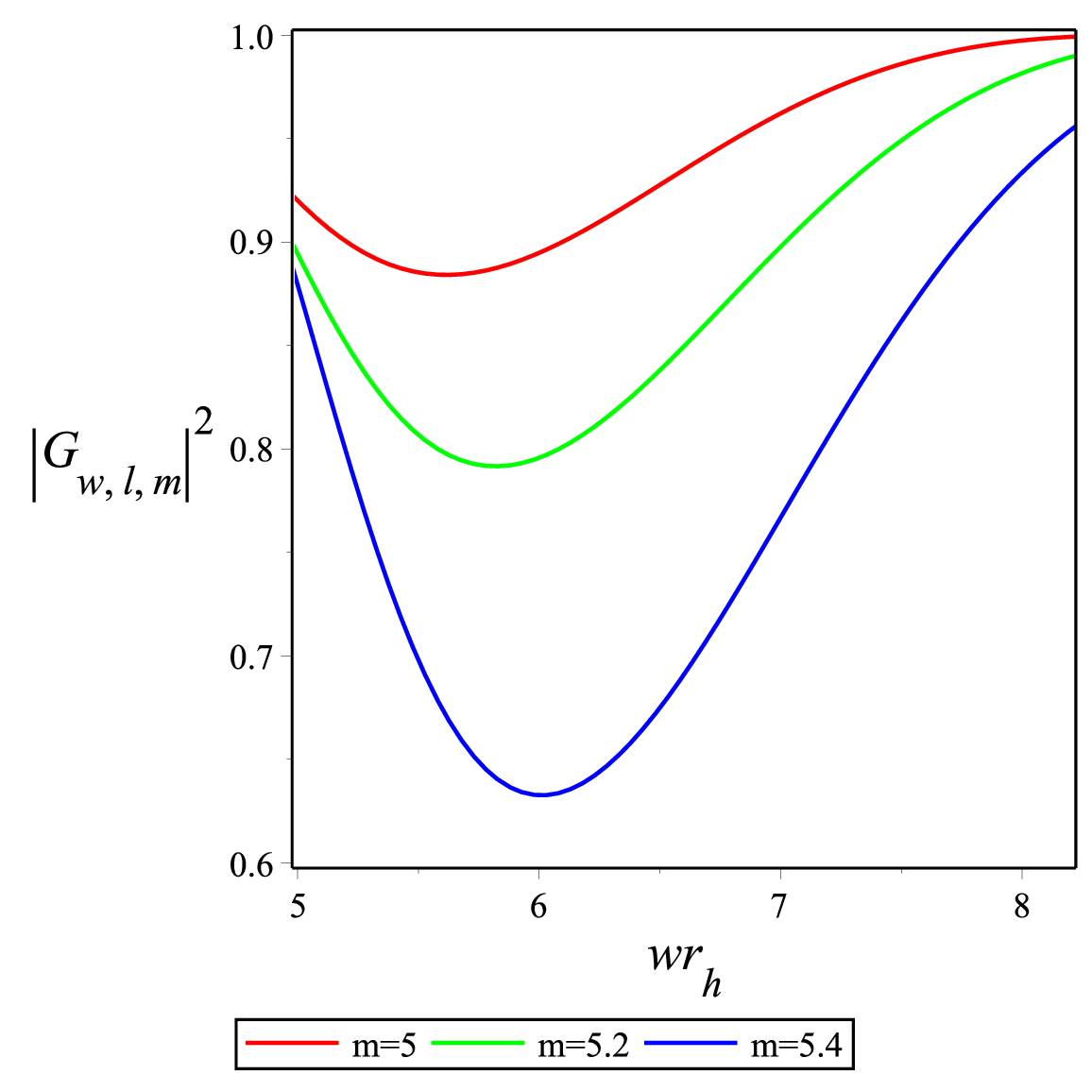,width=0.5\linewidth}
\epsfig{file=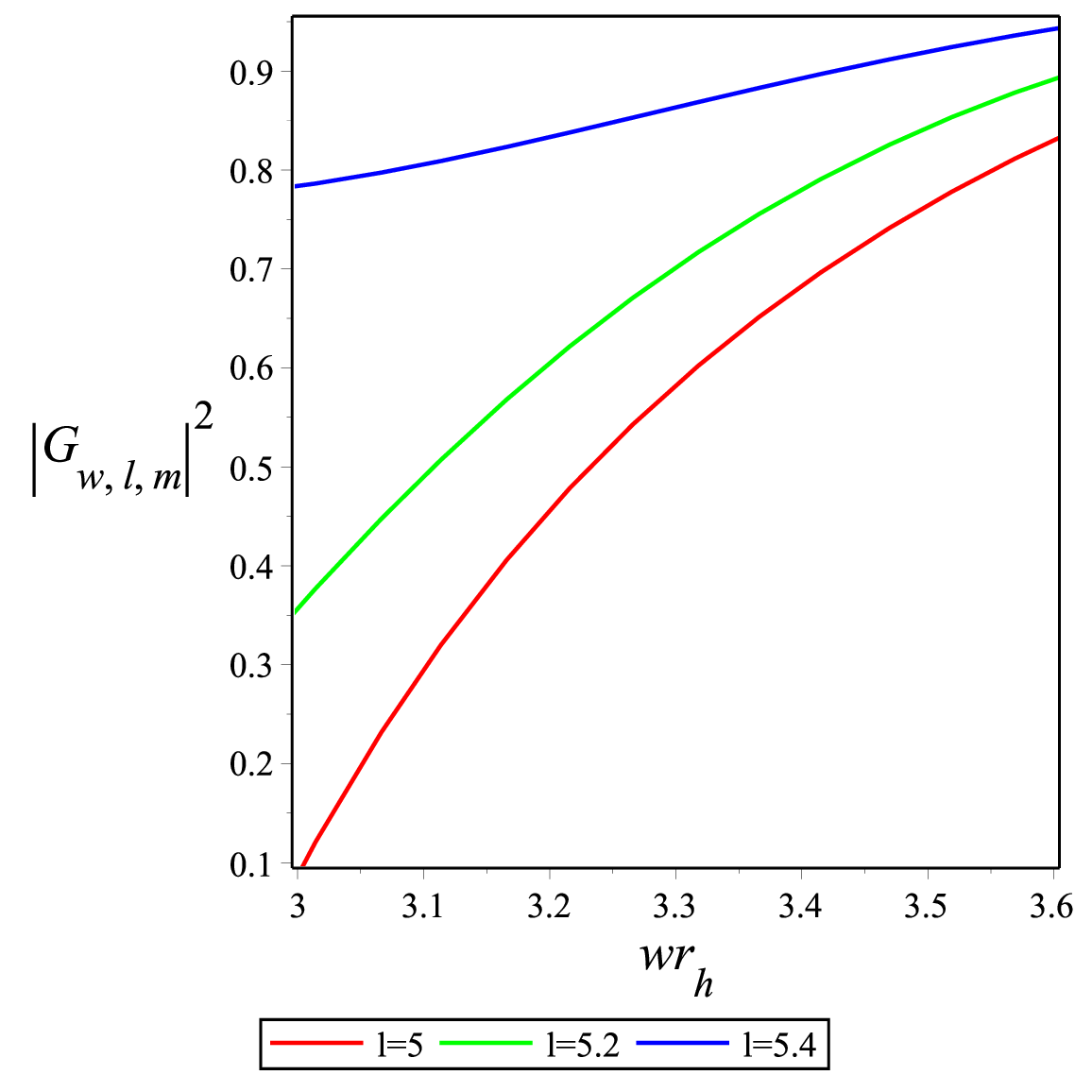,width=0.5\linewidth}\caption{GBF versus $wr_{h}$
with $l=5$ (left) and $m=5$ (right) for $q_m=3$, $q_e=3$, $p=1$,
$\alpha=1$, $\gamma=2$ and $r=1$.}
\end{figure}
\begin{figure}
\epsfig{file=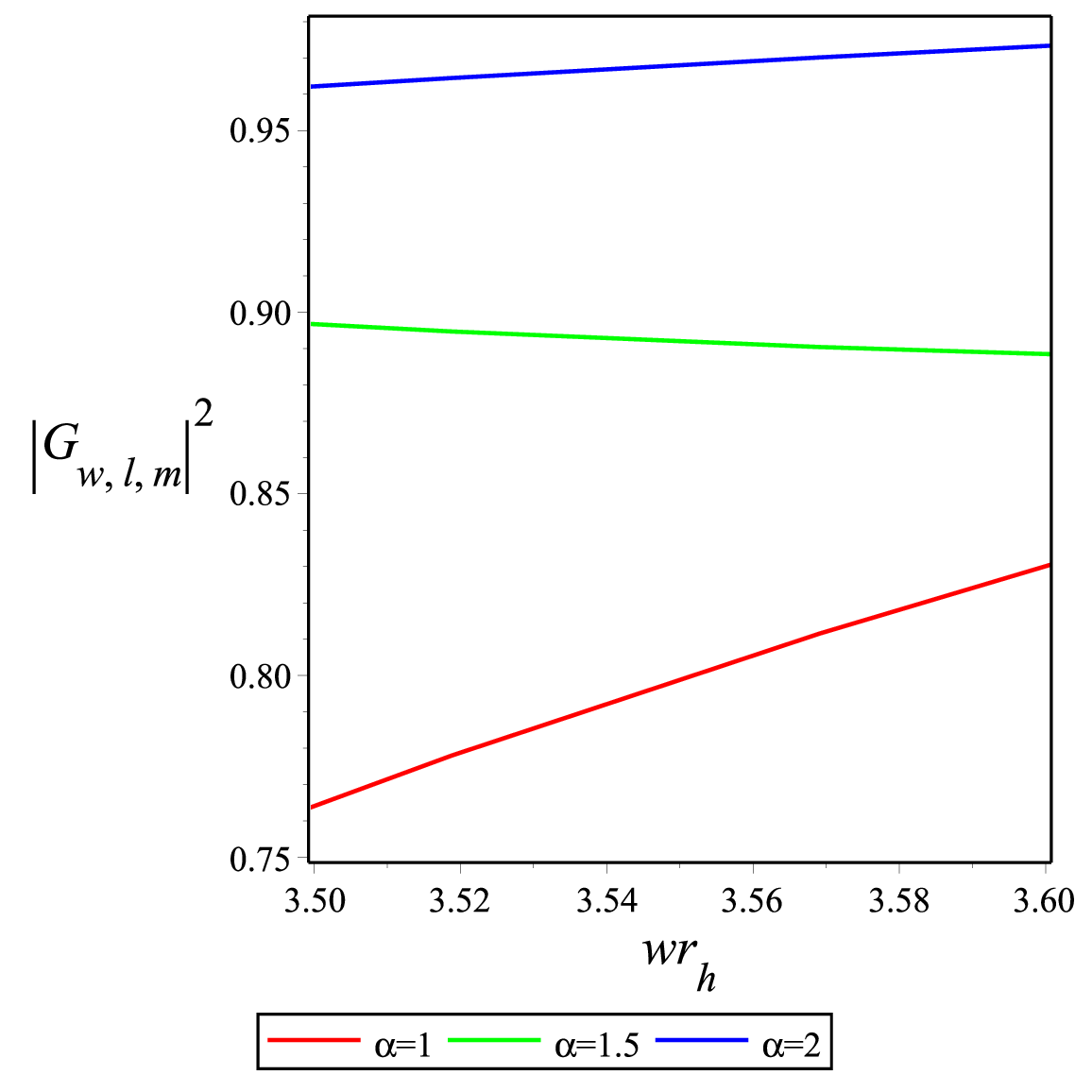,width=0.5\linewidth}
\epsfig{file=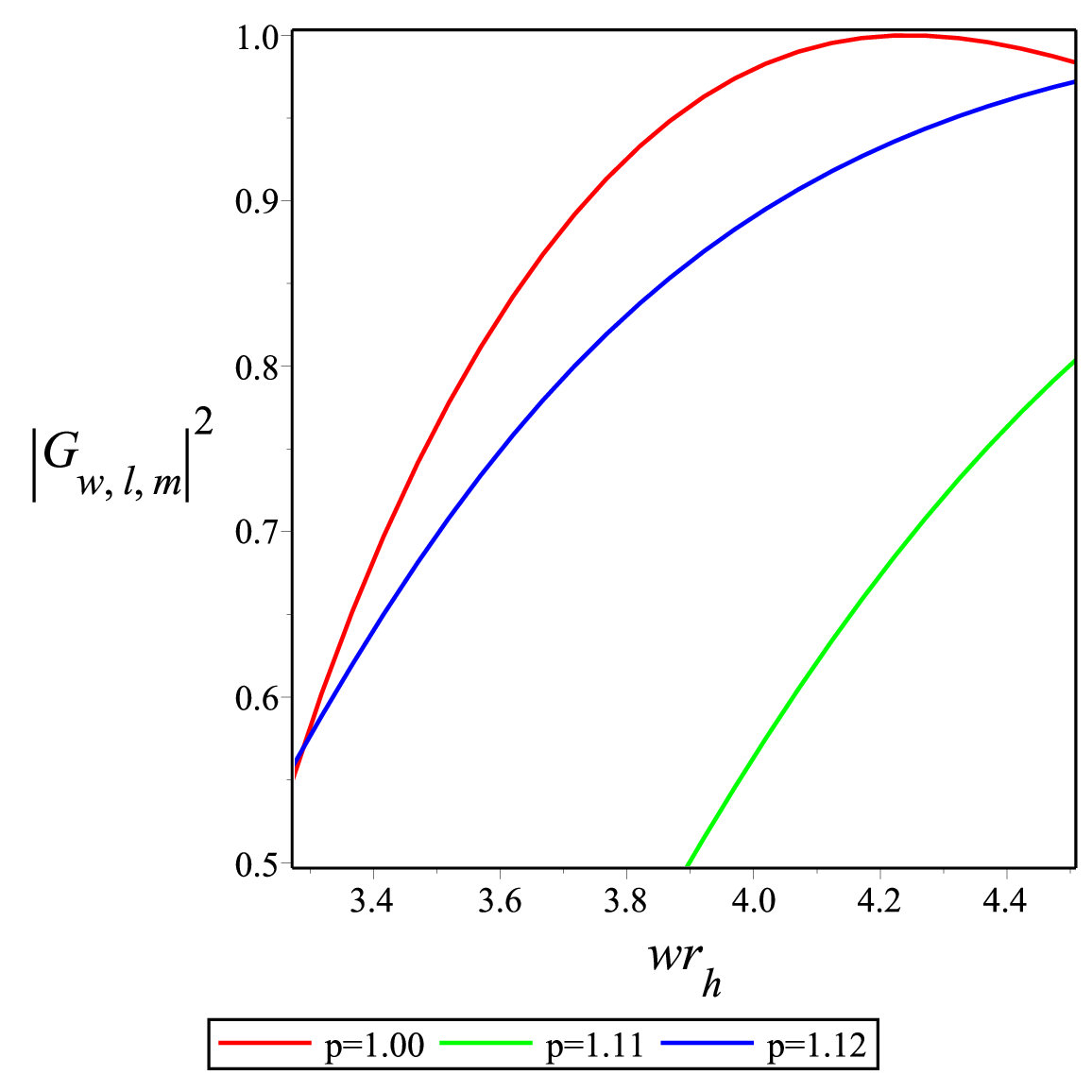,width=0.5\linewidth}\caption{GBF versus $wr_{h}$
with $p=1$ (left) and $\alpha=1$ (right) for $m=5$, $l=5$, $q_m=3$,
$q_e=3$, $\gamma=2$ and $r=1$.}
\end{figure}

We plot the graphs of GBF against frequency parameter for $m$, $l$,
$q_m$, $q_e$, $p$, $\alpha$, $r$ and $\gamma$. Figure \textbf{5}
shows the relation of GBF with the electric and magnetic charges,
i.e., the GBF increases with an increase in electric charge (left
panel) and magnetic charge (right panel). This indicates that GBF
increases for greater values of electric and magnetic charges which
shows that the BH with electromagnetic charge evaporates quickly.
Figure \textbf{6} indicates that the GBF decreases with an increase
in mass of the BH and increases for higher values of the coupling
parameter. This shows that the BH with greater mass and lower
angular momentum has a lower absorption probability. The graphical
behavior of $\alpha$ and AdS radius is shown in Figure \textbf{7}
which indicates that the GBF is directly proportional to $\alpha$
and inversely proportional to AdS radius. It shows that the decrease
in $\alpha$ and increase in AdS radius reduce the evaporation
process. In Figure \textbf{8}, we observe direct relation of the GBF
with radial coordinate which indicates that the BH of larger radius
with modified charge in anti-de sitter regime will die sooner.

In the RN AdS BH solution, the charge is modified by a parameter
$\gamma$ whose effects on GBF is discussed as follows.
\begin{itemize}
\item \textbf{Case 1: $\gamma>0$}
The graph of GBF is given in the left plot of Figure \textbf{9} for
subluminal case (the speed of modified electromagnetic radiations is
less than the speed of light). This indicates that BH evaporates in
the presence of non-linear electromagnetic modified charge and
greybody factor has an inverse relation with $\gamma$. Thus, higher
frequency of the Hawking radiations of a BH with electromagnetic
radiations having speed less than the speed of light reduces the
greybody factor, i.e., evaporation rate.
\item \textbf{Case 2: $\gamma<0$}
Here, we consider negative values of $\gamma$ and plot graph to
analyze the behavior of greybody factor. The right plot of Figure
\textbf{8} shows that BH does not evaporate when the speed of
modified electromagnetic radiations is greater than the speed of
light. This implies that higher frequency of the Hawking radiations
of a BH with electromagnetic radiations having speed greater than
the speed of light does not cross the gravitational barrier and BH
does not die i.e., evaporation rate becomes zero.
\end{itemize}
\begin{figure}\center
\epsfig{file=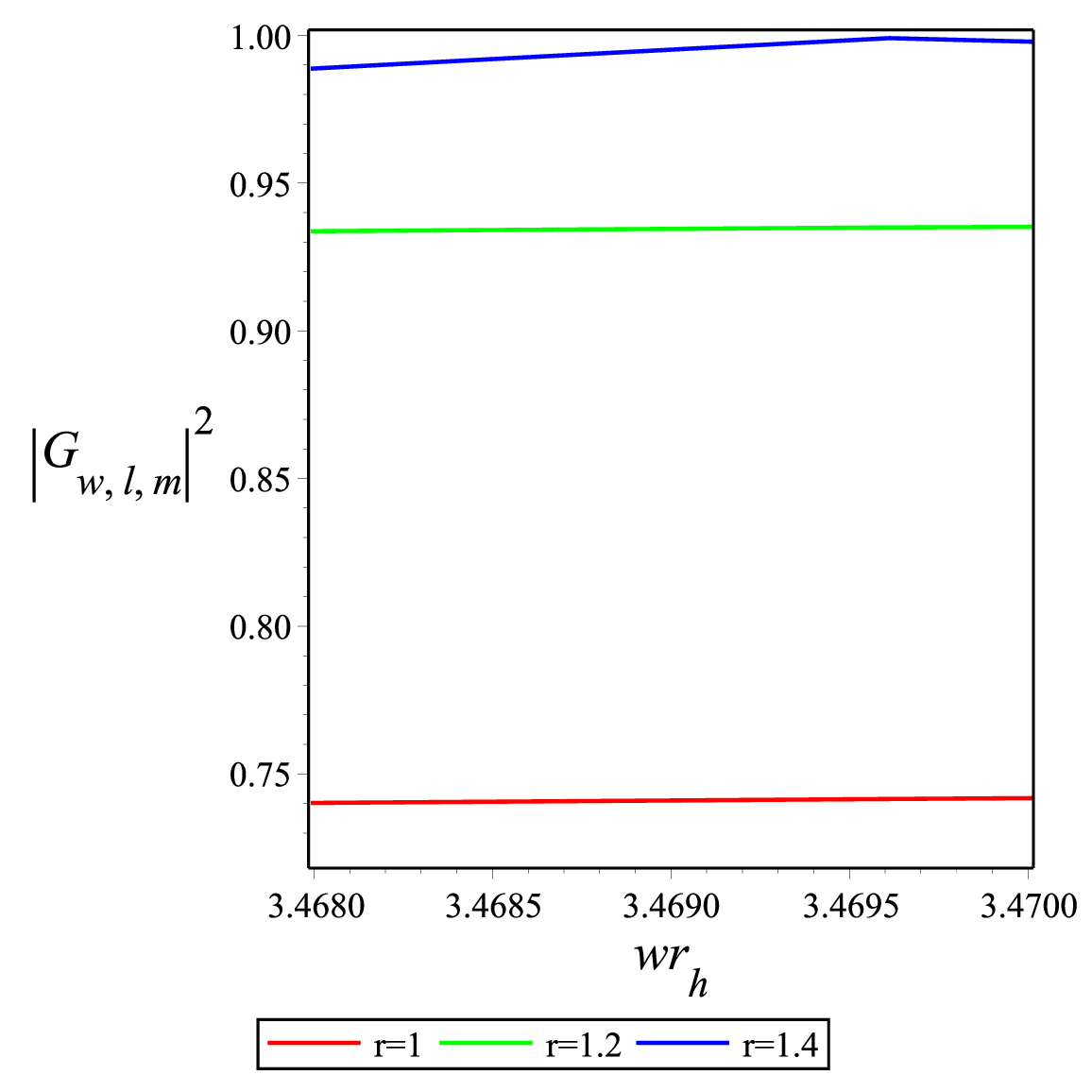,width=0.5\linewidth}\caption{GBF versus $wr_{h}$
with $\gamma=2$ (left) and $r=1$ (right) for $m=5$, $l=5$, $p=1$,
$\alpha=1$, $q_m=3$ and $q_e=3$.}
\end{figure}
\begin{figure}
\epsfig{file=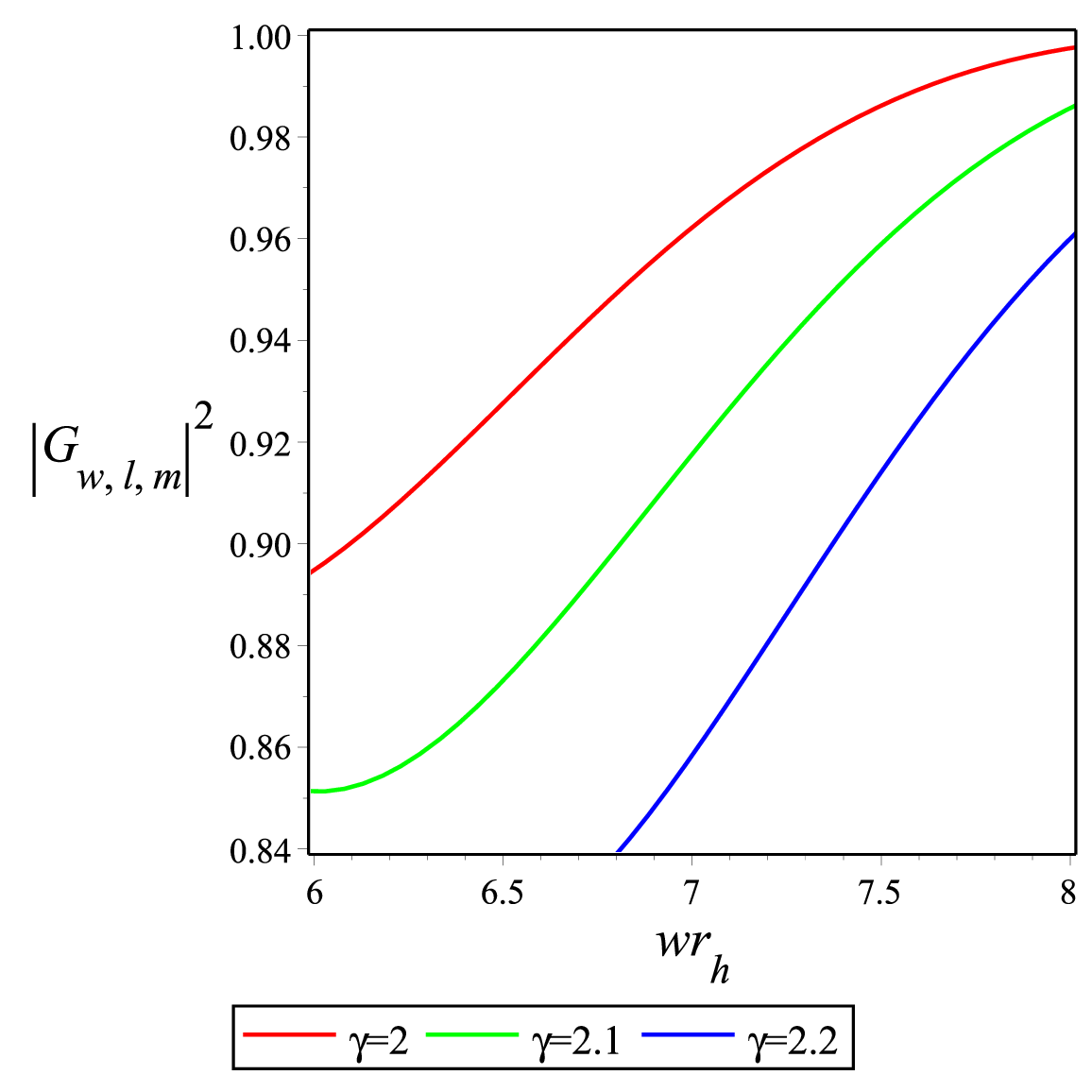,width=0.5\linewidth}
\epsfig{file=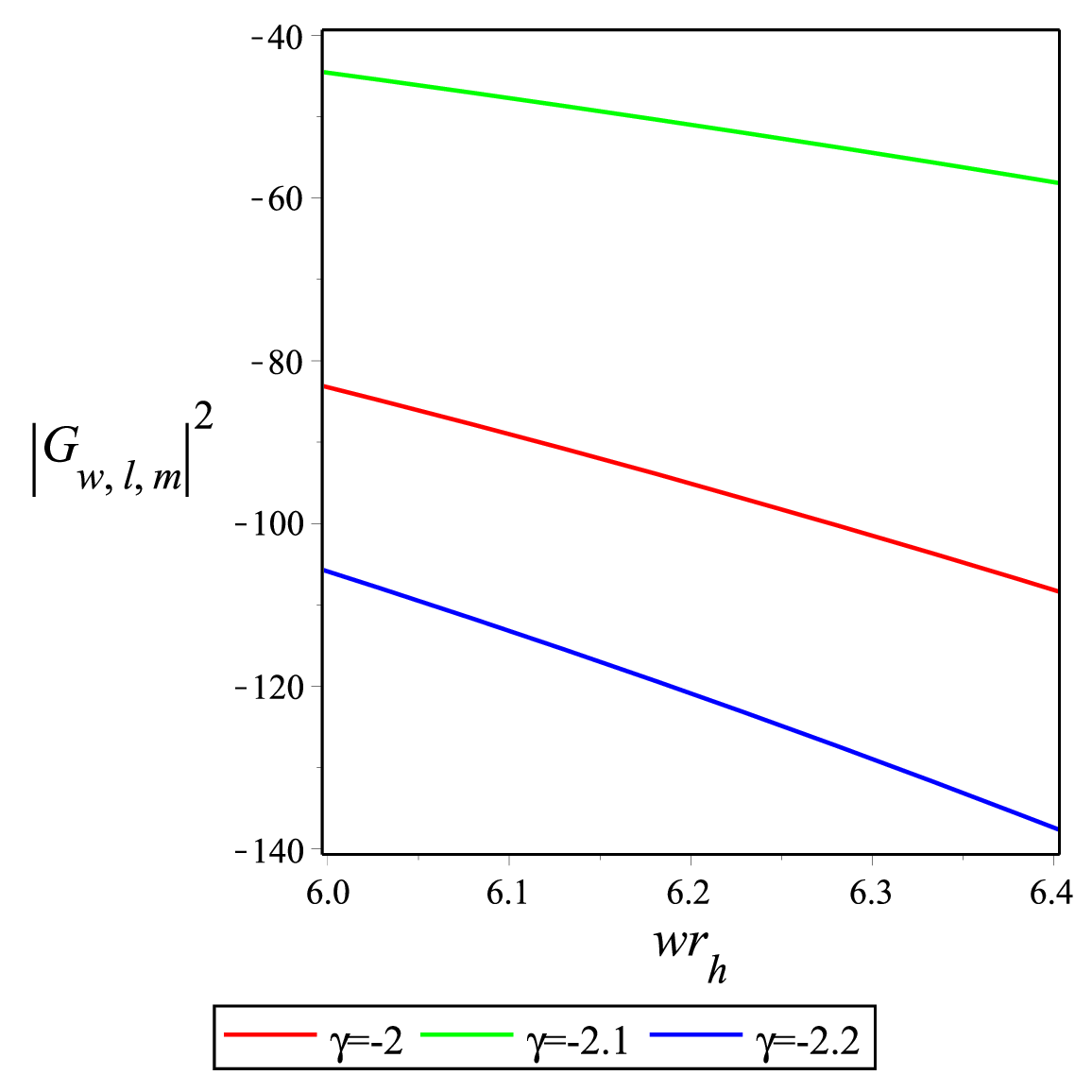,width=0.5\linewidth}\caption{GBF versus $wr_{h}$
with $\gamma>0$ (left) and $\gamma<0$ (right) for $m=5$, $l=5$,
$q_m=3$, $q_e=3$, $\alpha=1$, $p=1$ and $r=1$.}
\end{figure}

\section{Final Remarks}

In this paper, we have constructed an expression of the GBF for the
non accelerated charged modified BH in the AdS region. For this
purpose, we have computed a radial equation for the massless scalar
field with the help of Klein-Gordon equation. We have then
formulated the effective potential by transforming this equation
into Schrodinger wave equation. We have analyzed the graphical
behavior of effective potential for different values of various
parameters against $\frac{r}{r_h}$. We have used radial equation to
compute exact solutions in the form of HG function at event and
cosmological horizons. We have joined them in an intermediate regime
to enhance the feasibility of GBF over the entire domain. The main
findings of this paper are given as follows.
\begin{itemize}
\item The gravitational potential increases with an increase in electromagnetic charge (Figure
\textbf{1}), angular momentum (right panel in Figure \textbf{2}),
$\alpha$ and $w$ (left plots of Figure \textbf{3-4}). This shows
that the modified Maxwell theory reduces the absorption probability
of the BH and consequently minimizes the evaporation process.
\item The gravitational barrier decreases with the increase in mass (left panel in Figure
\textbf{2}), AdS radius (right panel of Figure \textbf{3}) and
$\gamma$ (right panel in Figure \textbf{4}) that increases the
emission rate of modified AdS charged BH.
\item The height of the GBF increases with an increase in electromagnetic
charges (Figure \textbf{5}), angular momentum (right plot of Figure
\textbf{6}), $\alpha$ and radial coordinates (left panels of
Figure(\textbf{7-8})). This indicates that GBF increases for greater
values of angular momentum, electric and magnetic charges.
\item The GBF decreases with an increase in mass (left panel of Figure \textbf{6}) and AdS radius
(right plots in Figure \textbf{7}). This shows that the emission
rate of BH becomes slow for higher values of $m$ and $p$.
\item The height of the GBF decreases with an increase in $\gamma$ (left plot of Figure \textbf{9}) for
subluminal case ($\gamma>0$). This indicates that the GBF and
evaporation rate of a BH with electromagnetic radiations having
speed less than speed of light decreases. The right plot of Figure
\textbf{9} shows that the BH with electromagnetic radiations having
speed less than the speed of light has no absorption probability.
\end{itemize}

Lai et al \cite{46} studied the GBF of the charged AdS BH with Maxwell
electrodynamics. We have extended this work for modified charged AdS
BH and discussed the effects of physical parameters on GBF. The GBF
for a regular de Sitter BH coupled (minimally/non-minimally) with
non-linear charge is analyzed in \cite{34}. We have found that the
effective potential of the BH increases with an increase in both the
mass and electric charge parameters ($q_m$ and $q_e$). This increase
in the effective potential indicates a reduction in the absorption
probability of the BH in an AdS spacetime. In the de Sitter regime,
there is an inverse relationship between the potential barrier and
the charge parameter. This implies that the charge parameter affects
the potential barrier differently in the de Sitter spacetime
compared to other scenarios. In the AdS region, angular momentum
enhances the emission rate of the BH, while in the de Sitter region,
it minimizes the emission rate. For small values of the cosmological
constant, the solution for the GBF in de Sitter spacetime remains
appropriate even if the distance between the BH horizons is large.
Conversely, larger values of the cosmological constant in AdS are
appropriate for smaller values of the radial coordinate. The
modification under consideration results in a reduction of the BHs
emission rate and an increase in its lifespan when compared to the
emission behavior in de Sitter spacetime. It is found that the
modification reduces the emission rate of the BH and increases its
life span as compared to the de Sitter spacetime. The GBF can be
used to measure the evaporation rate of a BH as it is the absorption
of radiations passing through the gravitational barrier. It would be
interesting to find the GBF of a
modified charged de Sitter and accelerating charged AdS BH.\\
\textbf{Data Availability:} No data was used for the research
described in this paper.

\end{document}